\documentclass[]{mn2e}  
\usepackage{graphicx}  
\usepackage{mnras_cite}  
\usepackage{color}
\definecolor{gris}{gray}{0.50}
\usepackage{color}
    
\newcommand{\nc}{\newcommand}

\nc{\beq}[1]{\begin{equation}\label{#1}}
\nc{\bea}[1]{\begin{eqnarray}\label{#1}}
\nc{\eea}{\end{eqnarray}}
\nc{\eeq}{\end{equation}}

\nc{\vev}[1]{\langle #1 \rangle}
\nc{\hyper}{\,\; F_{1{\hskip-16pt}2}{\hskip 11pt}}

\nc{\ie}{{\it i.e.}}
\nc{\eg}{{\it e.g.}}

\begin{document}  
    
\title[Local Bias]{
The Local Bias Model in the Large Scale Halo Distribution}
    
\author[Manera \& Gazta\~{n}aga]{M.Manera$^{1,2}$ \&  E.Gazta\~{n}aga$^{3}$\\  
$^{1}$Institute of Cosmology and Gravitation, University of Portsmouth, Dennis Sciama Building, Burnaby Road, Portsmouth PO1 3FX, UK \\
$^{2}$Center for Cosmology and Particle Physics, New York University, 4 Washington Place, NY 1003, New York,USA\\
$^{3}$Institut de Ci\`encies de l'Espai, CSIC/IEEC, Campus UAB,
F. de Ci\`encies, Torre C5 par-2,  Barcelona 08193, Spain\\
}

\maketitle  

\begin{abstract}     
We explore the biasing in the clustering statistics of halos as
compared to dark matter (DM) in simulations.  We look at the second
and third order statistics at large scales of the (intermediate)
MICEL1536 simulation
 and also measure directly 
the local bias relation $h = f(\delta)$ between DM fluctuations,
$\delta$, smoothed over a top-hat radius $R_s$ at a point in the
simulation and its corresponding tracer $h$ (i.e. halos)
at the same point. This local relation can be Taylor
expanded to define a linear ($b_1$) and non-linear ($b_2$)
bias parameters.  The values of $b_1$ and $b_2$ in the simulation
vary with $R_s$ approaching
a constant value around $R_s >30-60$ Mpc/h.
We use the local relation  to predict the clustering of the tracer in terms of the one
of DM.  This prediction works very well (about percent level) 
for the halo 2-point correlation $\xi(r_{12})$ for $r_{12}>15$ Mpc/h, but only
when we use the biasing values that we found at very large smoothing radius 
$R_s >30-60$ Mpc/h.
We find no effect from stochastic or next to leading
order terms in the $f(\delta)$ expansion. But we do find some
discrepancies in the 3-point function that needs further
understanding. 
 We also look at the clustering of the smoothed
moments, the variance and skewness which are volume average
correlations and therefore include clustering from smaller scales.  In
this case, we find that both next to leading order and discreetness
corrections (to the local model) are needed at the $10-20\%$
level. Shot-noise can be corrected with a term $\sigma_e^2/\bar{n}$
where $\sigma_e^2<1$, i.e., always smaller than the Poisson
correction.  We also compare these results with the peak-background
split predictions from the measured halo mass function. We find 5-10\%
systematic (and similar statistical) errors in the mass estimation
when we use the halo model biasing predictions to calibrate the mass.

\end{abstract}    
    
    
\maketitle    
    
    
\section{Introduction}    
\label{sec:intro}    

To do precision cosmology it is important to understand
accurately galaxy bias, i.e., 
how the spatial distribution of galaxies is related to the underlying dark matter distribution. 
Because galaxies are known to form in dark matter halos, its biasing
can be approached in two natural steps. The first step is the bias between halos and dark matter. 
The second step is the bias between galaxies and halos, which is commonly approached by means of models 
of galaxy occupation in halos (see for instance Zheng et al 2005 \& 2009, Brown 2008, Tinker 2006 \& 2010).

Biasing requires a complex modeling and in this paper we will 
focus on  the first step only. This means that our findings might not
be directly applicable to galaxy surveys. In the limit in which halo
biasing resembles galaxy biasing or in the limit where observations are
good tracers of the halo distribution (i.e., for galaxy groups or
clusters) our results will be of direct relevance to the
interpretation  of clustering statistics in galaxy and cluster surveys.

We will study the halo bias in a big cosmological 
dark matter simulation from the MICE collaboration (Fosalba et al. 2008,
Crocce et al. 2009). 
\footnote{For more information about the MICE collaboration team and
  the simulations, see  http://ice.cat.es/mice/}.
We will address two main questions:
a) how accurate is the so-called \textit{local} bias model
to predict clustering statistics, and  b) how bias predictions 
from the mass function compare with the ones in local model.
 In the process of answering 
these questions we will also learn about nonlinear and stochasticity contributions to the halo variance.   
The local bias model, introduced by Fry and Gazta\~naga (1993), assumes 
a general non-linear (but local and deterministic) relation between the smoothed density contrast 
in the distribution of halos (or galaxies) 
and the smoothed density contrast of the dark matter, i.e., $\delta_h = F[\delta_m]$.
In reality, bias is stochastic and not quite deterministic
(eg., see Somerville et al. 1999, Tegmark \& Bomley 1999, Dekel \& Lahav 1999) 
and at some level, due to tidal forces and evolution, it will have non-local and
anisotropic contributions.  It is also not clear to what extent the
halo  or galaxy density should depend only on the underlying matter density, 
without including other direct dependencies (like the gravitational
potential or velocity fields, for instance). Bias could also relate to
mass at some initial condition or in Lagrangian space (see,
Catelan, Matarrese \& Porciani 1998, Matsubara 2008 and references therein).

The bias parameters of the local model are the coefficients of the Taylor expansion of $F[\delta]$, and depend
on the halo and dark matter smoothing scale. In this paper we show that for halo samples with a minimum
mass less than $10^{14}$ solar masses the bias parameters converge at a smoothing scale $\sim 30$-$60$ Mpc/h. 
We can then compare these local bias parameters obtained by directly fitting $F[\delta]$ in the simulations
with the bias from clustering measurements like the two and 3-point correlations functions, the variance
and the skewness. We will show that the local bias model
works well at least within a few percent level. 
When this local model is applied to interpret real galaxy surveys it
can be used to recover information about dark matter clustering
and biasing parameters.

One common way to predict the bias parameters is to use the peak
background split Ansatz.  Bias parameters are predicted from the mass
function using few assumptions: locality and also the assumption that
the conditional mass function of an overdense (underdense) patch of
the universe can be treated as if it were equal to the average mass
function of the universe at a different time, or mean density. Peak
background split predictions for the bias, specially from the Sheth
Tormen mass function (Sheth and Tormen 1999), and the Press-Schechter
mass function (Press \& Schechter 1974, PS from now on) have been used
a lot in the literature.  In the second part of this paper we will
compare these bias predictions with the bias from clustering, and with
the local bias, and study their dependence on the halo mass threshold
used to fit the mass function. The inaccuracy of the peak background
split has also been studied in Manera, Sheth and Scoccimarro (2010)
with complementary results to those of this work.

The peak background split bias parameters predicted from the
Press-Schechter (PS) mass function \textit{together} with the
assumption of the local bias was tested in a precursor paper by Mo,
Jing and White (1997), where the local model was used to compare the
skewness and higher order moments of halos in N-Body simulations with
predictions and observations, leading to the conclusion that the
galaxies from the APM survey (as measured in Gaztanaga 1994) should
not be highly biased.  Mo, Jing and White used a small simulation of
only 256 Mpc/h and $128^3$ particles with plots that show no
errorbars.  In some of their plots, specially when halos are
identified at the same time that moments are calculated, differences
between theory and simulations could be interpreted as being
significant for our current precision requirements. Unfortunately
since they tested the PS bias parameters and the local bias
model together, it is unclear how each assumption contributes to the
mismatch.

In a follow-up paper, Casas-Miranda, Mo and Boener (2003) redid the
previous analysis, this time with the Sheth and Tormen mass function,
and applied the results to the Lyman break galaxies at $z=3$. Their
plots of skewness and higher order moments still show no error bars,
and differences between theory and simulations could amount more than
15\% in some cases.  Again, the question arises to whether the local
bias is a good approximation or not independent of the bias prediction
from the mass function, which requires extra assumptions and varies
depending which mass function one decides to use.  In our paper we can
separate these effects by obtaining the local bias parameters directly
from a fit of the local bias relation $F$, thus testing the local
model separately, from the bias predictions.  A failure of the local
bias model could point towards what other contributions should be
included next (if any) when analysing observational data to the
precision needed for the current generation of surveys.

Another difference in our analysis with respect to the previous works
above is that we study both moments (variance and skewness) and 2 and
3-point correlation functions. Moments are closer to
the local relation in that they are both smoothed (spherically
averaged)  quantities,  so one would expect better agreement for them. But
they suffer from shot-noise (or discreteness effects) and stronger
non-linear effects (as they include clustering on all scales
smaller than the smoothing radius). The 2 and 3-point
functions do not suffer from shot-noise and can better separate
the effect of different scales (because they are averaged over radial
shells rather than integrated over spheres). Moreover, the 3-point function provides
different information than the skewness. Both are related third order
statistics, but the 3-point function also gives shape
information (i.e., how elongated are the triangles) which is missing
in the skewness. We also study the 2-point cross-correlation of mass
and halos which gives an idea of how
important the stochasticity is in the bias relation.

The relation between the mass function and the bias can be inverted.
Consequently one may use the bias
as a proxy for the mass of the halo sample. 
This is of direct relevance to the interpretation of observational data.
Systematic errors in
estimating the mass from the bias would propagate to, and broaden, the 
constraints on cosmological parameters (like the dark energy equation of state parameter
$w$) when fitted to the estimated halo mass function. 
Notice that self calibration methods for the mass function,
which are expected to be used by DES-like surveys, assume that we know
the mass-bias relation  (Lima \& Hu 2005, 2007). In this work we will
assess how well the halo mass is recovered by using as input 
the clustering bias parameters from the two and from the 3-point correlation functions. 

Going one step further, to relate halos to galaxies,
it has become customary to use the Halo Occupation Distribution 
(HOD) prescription, which consists of populating either theoretically, 
or in the simulation, the dark matter halos 
with galaxies, using some simple (author-variable) population function 
that usually depends only on three or four parameters. 
The HOD prescriptions are far from providing a few percent precision of all measurements. 
For instance, Scoccimarro et al. (2001)
found that they were unable to match both the variance and the skewness of APM galaxies. 
Since a local model of biasing is assumed along with the bias prediction from the mass function, 
it is unclear if this disagreement is due to the 
HOD choice or to the failure of either the local model
or the bias predictions. It is therefore of direct
observational interest 
to assess each step separately, which is what this paper starts doing. 

Indications that more work is needed to construct reliable galaxy
mocks has been given by Guo and Jing (2009).
Using a semianalytical mock sample of galaxies constructed from an N-Body
simulation, they compared the local bias parameters from
clustering with the local bias predicted using the peak background
split Ansatz plus an HOD, which were found to be significantly
different. Such difference may arise from the fact that the authors
were using  a published prescription instead of 
fitting their own HOD function, but part of the disagreement could
come as well from the local bias and the PBS Ansatz.

  Finally, note that we only
study clustering in configuration space. Bias will most likely have
different effects in Fourier space, in particular regarding shot-noise
effects and stochasticity (see eg. Tinker et al. 2010, Seljak, Hamaus
\& Desjacques 2009, Cai, Bernstein \& Sheth 2010, and references
therein).

The paper is organized as follows. Section II gives 
a brief introduction to the simulations.
In section III we study the
performance of the local bias model in simulations and present
a study of shot-noise and next to leading order contributions.
In section IV we compare the clustering of halos in the simulations
with the peak backgrounds split predictions and check 
how well we can recover the mass of halos from the bias parameters. 
We present our conclusions in section V.

\section{Numerical Simulations}

In this paper we work with the comoving data from the MICE
intermediate dark matter simulation, which has a volume of $V= ( 1536
Mpc/h) ^3$ and $N = 1024^3$ particles, and consequently a mass
resolution of $2.34\; 10^{11} M_{\odot}/h$ This simulation have been
run with Marenostrum at the Barcelona Supercomputer Center using the
L-GADGET code, periodic boundary conditions, and 128 processors.  In
this paper we use the $z=0.0$ and the $z=0.5$ comoving outputs.
The cosmological model parameters for the simulation are
$\Omega_m=0.25$, $\Omega_\Lambda=0.75$, $\Omega_B=0.044$, $n_s=0.95$,
$h=0.7$ and $\sigma_8=0.8$. The softening length of the simulation is
50 Kpc/h. The initial conditions were set at $z=50$ using Zeldovich
approximation.  Halos have been found using a Friends of Friends
algorithm with a linking length 0.168 times the mean interparticle
distance, which results in 2729833 halos of more than 20 particles at
z=0, and 2110669 halos at z=0.5. The effect of chancing linking length
have been studied in Manera, Sheth and Scoccimarro 2010.  By working
with comoving data we concentrate on the gravitational evolution and
structure formation and get rid of redshift distortions and other
lightcone effects, which might not be directly related to the
questions addressed here. Nevertheless, since at the end we want to
model observational data, the inclusion of lightcone effects and
redshifts distortions have to be considered as the natural next step
in this study (see also Marin et al 2008).

\begin{figure*}
\vskip -0.5cm
\hskip -0.2cm
\includegraphics[width=84mm]{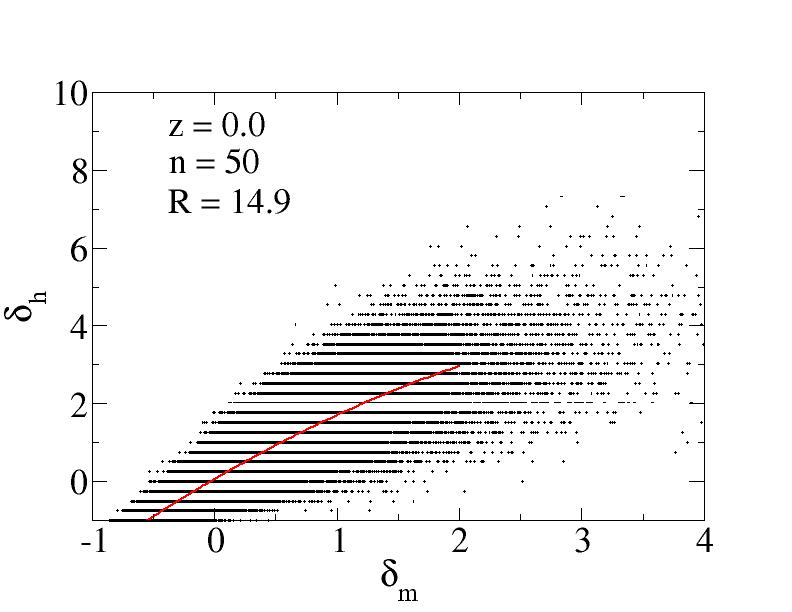}
\includegraphics[width=84mm]{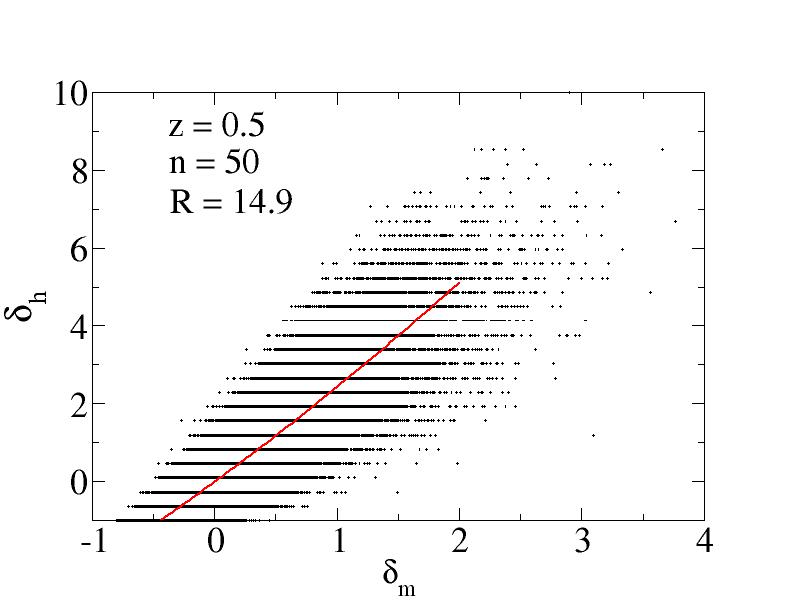}
\vskip -0.1cm
\hskip -0.2cm
\includegraphics[width=84mm]{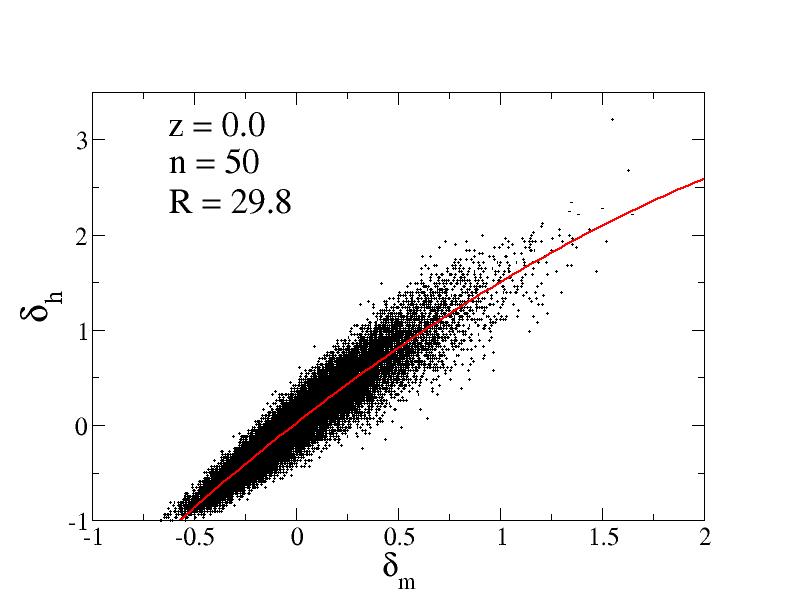}
\includegraphics[width=84mm]{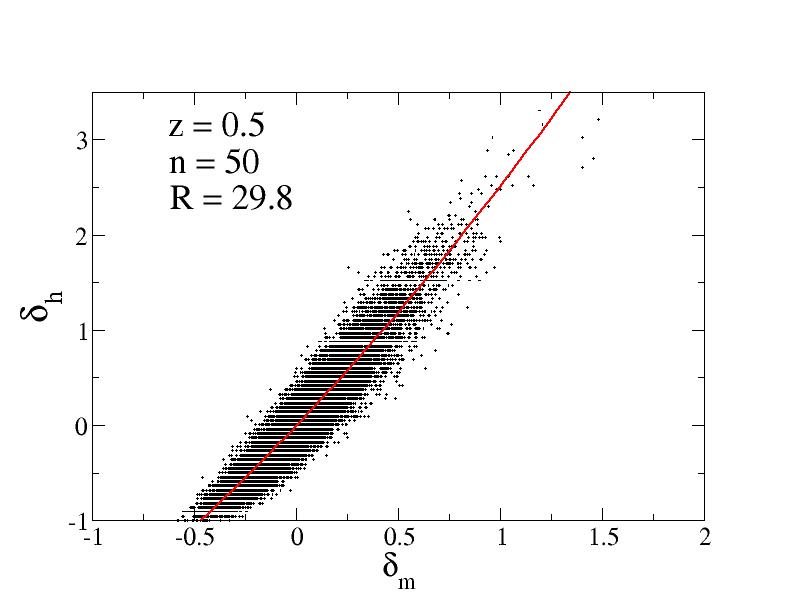}
\vskip -0.1cm
\hskip -0.2cm
\includegraphics[width=84mm]{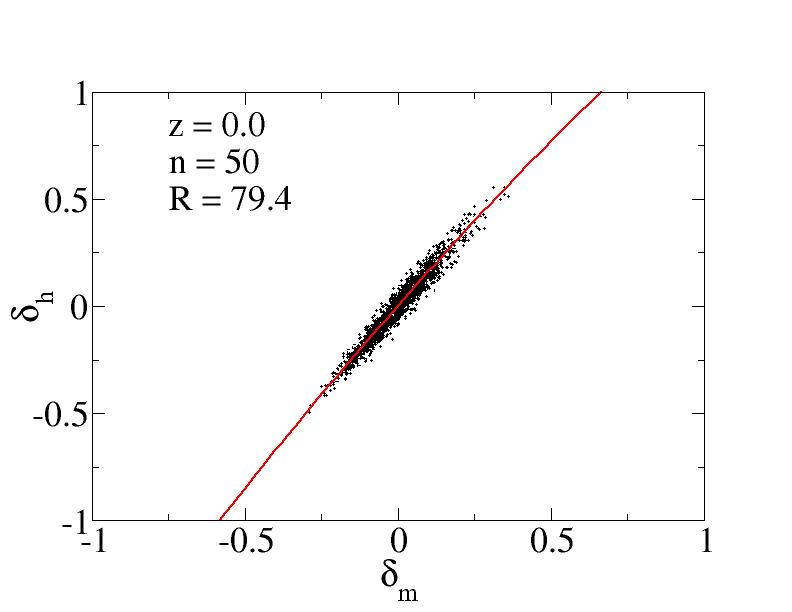}
\includegraphics[width=84mm]{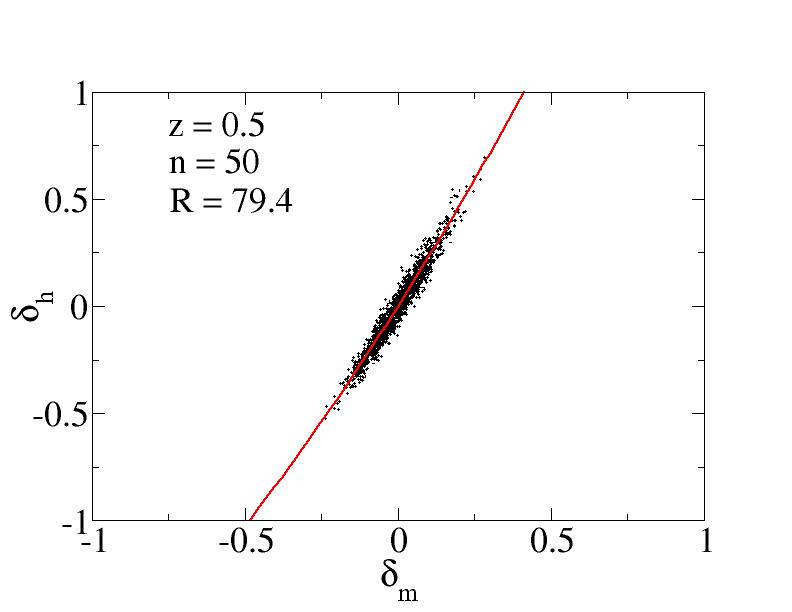}
\caption[Scatter ($\delta_h$,$\delta_m$) plots for different smoothing radius]
{Scatter plots showing halo density contrast $\delta_h$, 
smoothed over top-hat cells, for halos of 50 or more
particles versus dark matter density fluctuations  $\delta_m$ smoothed
over the same cells. Results are shown for a different cell sizes with equivalent radius $R_s$ as 
labeled in the figure. Results are for simulation data at redshift z=0
(left panels) and z=0.5 (right panels).  
In a continuous line we show the least square fit to the local bias parabola.}
\label{scatterzRs}
\end{figure*}

\begin{figure*}
\vskip -0.5cm
\center
\includegraphics[width=84mm]{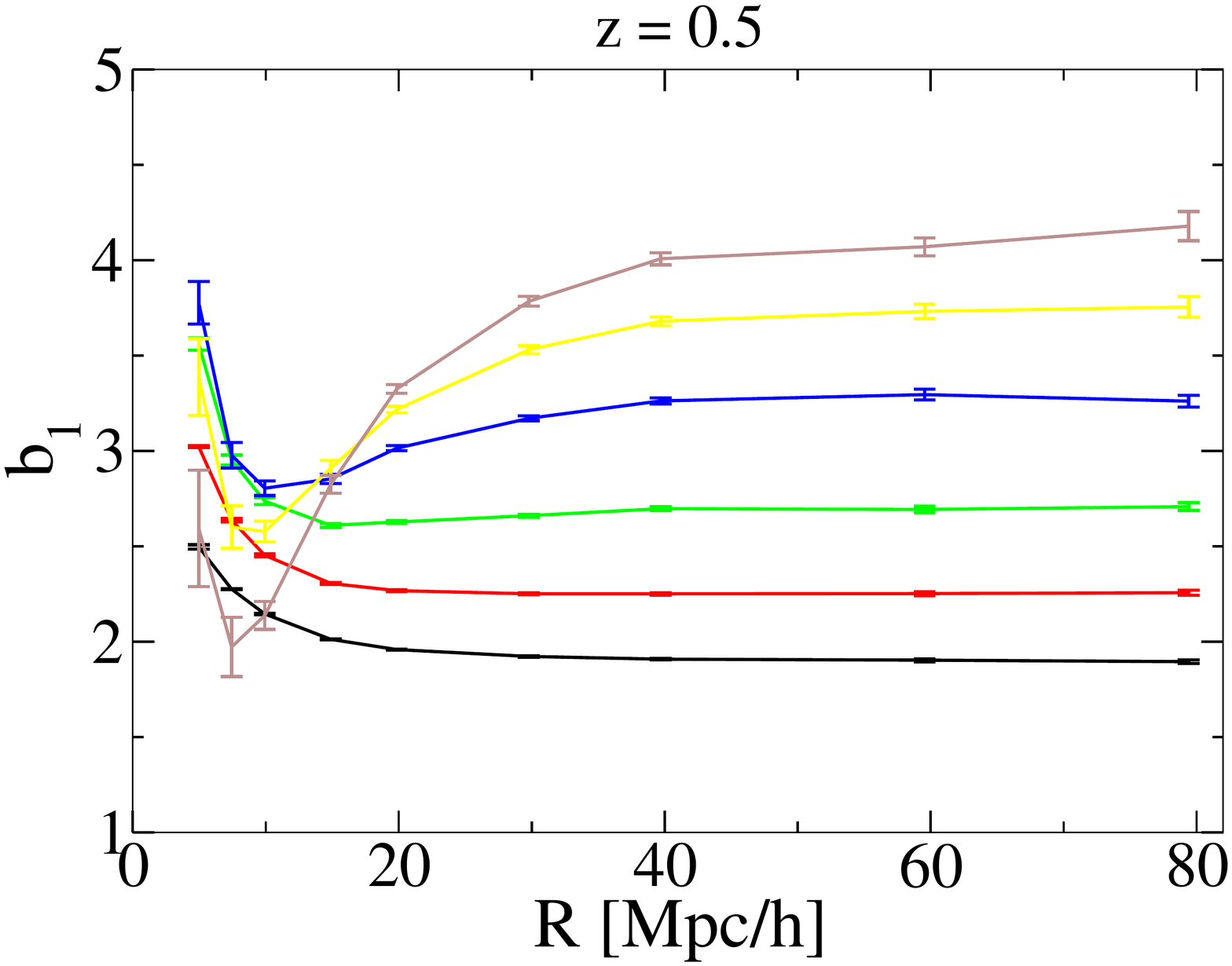}
\includegraphics[width=84mm]{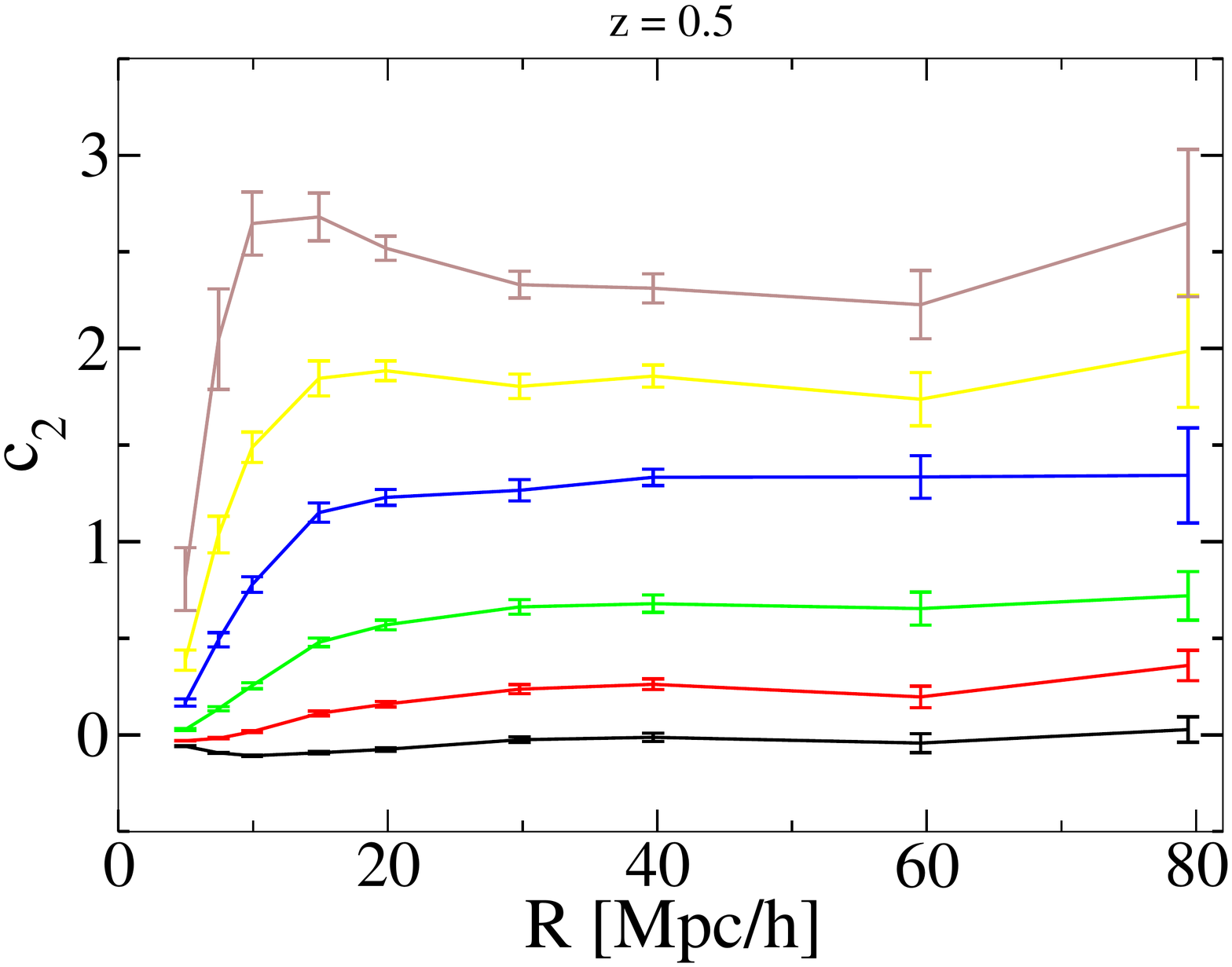}
\vskip -1.cm
\includegraphics[width=84mm]{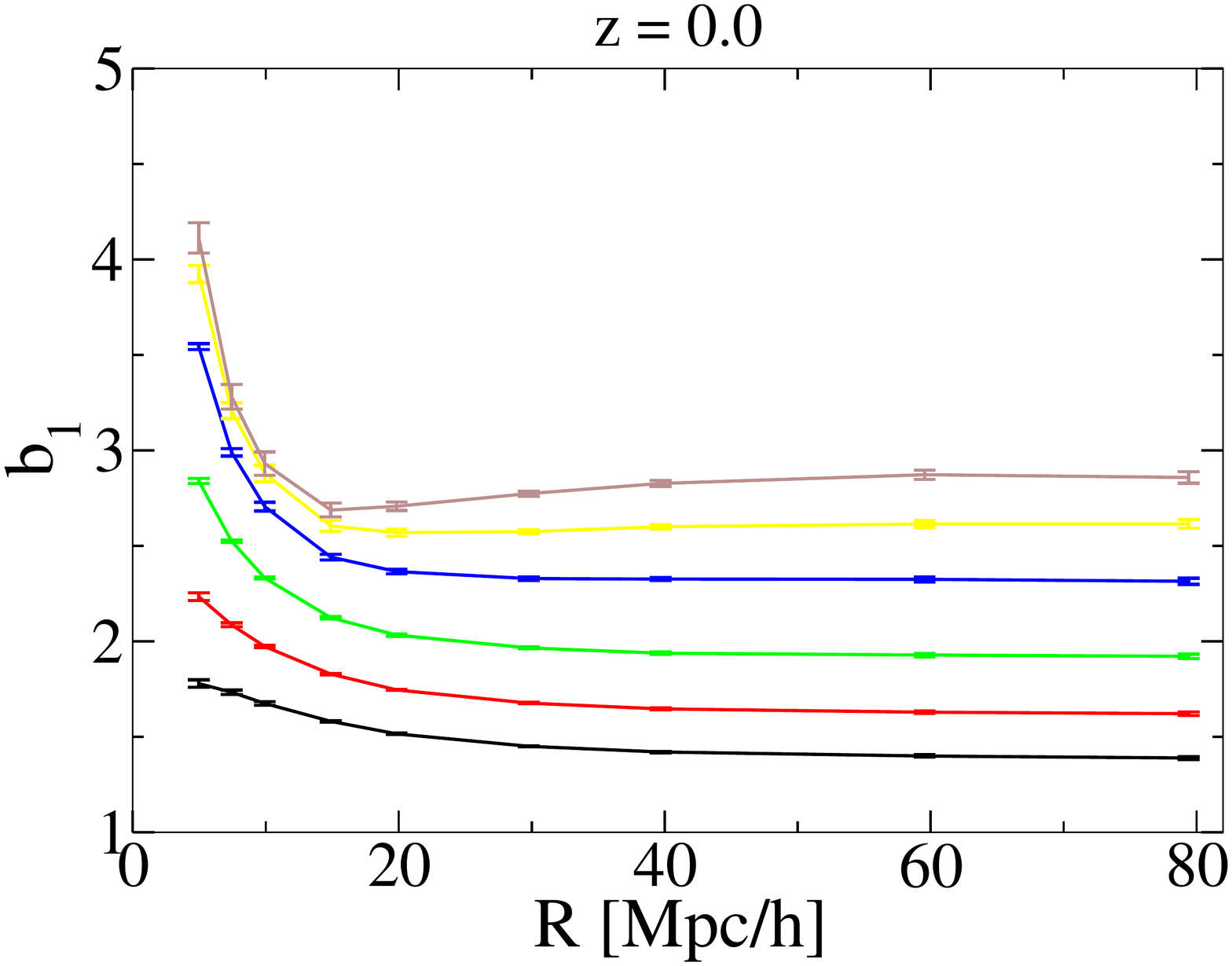}
\includegraphics[width=84mm]{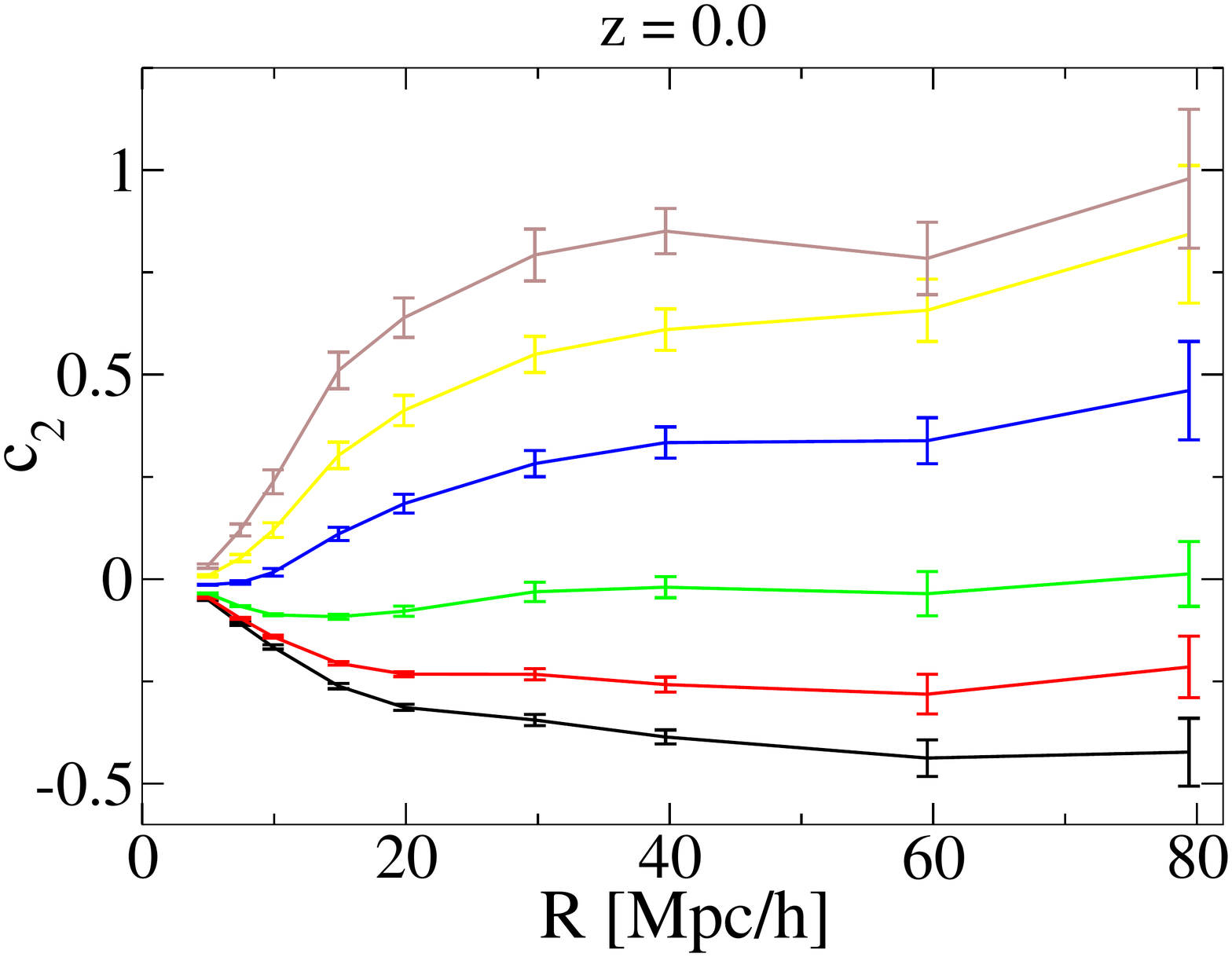}
\caption[Dependence of $b_1$ on the smoothing radius]
{Variation of $b_1$ (left panels) and $c_2$ (right panels)
as a function of the smoothing radius $R$. Top panels
corresponds to $z=0.5$ and bottom panels to $z=0$. Results are shown
for different minimum number of particles per halo, $n$. 
In each panel from bottom to top n=25 (black), n=50 (red), n=100 (green), n=200 (blue) and n=400 (yellow). Being the particle mass $23.42 \times 10^{10} M_{\sun}$ it yields, after correcting for resolution effects, minimum masses of 0.5 (black), 1.06 (red), 2.19 (green), 4.49 (blue), and 9.11 (yellow) $10^{12} M_{\sun}$} 
\label{figb1Rs}
\end{figure*}

\section{Local Bias Performance}    
\label{sec:localbias}    

A simple model for halo or galaxy bias was introduced by Fry and Gazta\~naga (1993).
These authors assumed that the density contrast in the halo (or more
generally in the galaxy) distribution
$\delta_h$ can be expressed as a non-linear function of the local density
contrast of dark matter, $\delta_m$, so that $\delta_h=F[\delta_m]$. On large enough
smoothing scales, where the fluctuations are small, this relation can be
expanded in a Taylor series
\beq{deltataylor}
\delta_h=F[\delta_m] =
\sum_{k=0}^\infty \frac{b_k}{k!}\delta^k_m=b_0+b_1\delta_m+\frac{b_2}{2}\delta^2_m+\cdots 
\eeq
where $\delta_m(r)$ is the local density contrast at position $r$ 
smoothed over a given characteristic $R_s$ scale.
With this local bias model, one can compute the 2 and 3-point halo biased correlation functions, 
to find \cite{Fry93,Frieman94} 
\bea{Q3G}
\xi^h(r_{12}) &\equiv& <\delta_h(r_1)\delta_h(r_2)>  ~
\simeq ~b_1^2 ~\xi(r_{12}) \nonumber\\
Q_3^h(r_{12},r_{23},r_{13}) &\simeq& {1\over{b_1}} ~\left[ Q_3(r_{12},r_{23},r_{31}) +c_2 \right]
\eea
where  $c_2 \equiv b_2/b_1$ and
$\xi(r_{12}) \equiv <\delta_m(r_1)\delta_m(r_2)>$
is the 2-point matter correlation function. The distance $r_{12}$ corresponds to the separation of two
arbitrary positions 1 and 2. The hierarchical 3-point function, $Q_3$ has 3 such distances and 
is defined as
\beq{Qdef}
Q_3(\alpha) \equiv \frac{<\delta(r_1)\delta(r_2)\delta(r_3)>}{
{ [ \; \xi (r_{12})\xi(r_{23})+\xi (r_{12})\xi(r_{13})+\xi (r_{13})\xi(r_{23})  \; ] }}  
\eeq
We need three parameters to specify the triangle formed by the 3
positions $r_1$, $r_2$ and $r_3$. We will fix two of these sizes
($r_{12}$ and $r_{13}$) and
show the results as a function of $\alpha$, the angle between
$r_{12}$ and $r_{13}$. In general $Q_3(\alpha)$ has a characteristic
U or V-shape (see Fig.\ref{Q3z}): is larger for small and large angles than
for intermediate values.

There is ambiguity over what we should use as smoothing scale $R_s$
in Eq. \ref{deltataylor}. A common and natural choice is that $R_s$ should be smaller than
$r_{12}$. But we will show that this does not provide a good model for $r_{12}< 60$ Mpc/h. 
 Another possibility, that we will support here,  is to think of an effective $R_s$ that can be  
 larger than $r_{12}$. Correlations are estimated from spatial
 averages over very large volumes and one can then think of Eq.\ref{deltataylor}
as some average transformation over the whole volume. In this sense
this bias transformation just provides an effective description which
is only local over a very large smoothing radius. 

When the correlation distances are zero we recover the corresponding
relations  between smoothed variances and skewness (see Eq.\ref{varskew}).
\beq{eq:S3h} 
\sigma^2_h \simeq b_1^2 \sigma^2  ~~;~~  S_3^h \simeq (S_3 + 3 c_2 )/b_1 
\eeq
Here it is common to identify $R_s$ with the smoothing scale in variance
and skewness, but this does not need to be the case.

In the above equations, and also in Eq.\ref{Q3G}, the $\simeq$ sign indicates that this
is the leading order in $\xi$. 
Note that in general, even when $\delta_m \ll 1$, the linear bias prescription (i.e., using only $b_1$) 
is not accurate for higher-order moments like $Q_3$, the reason being that nonlinearities in bias (i.e. $c_2$)
generate non-Gaussianities of the same order as those of gravitational
origin. In general, to predict higher order correlations in halo (or
galaxies) to order $N$ the local relation has to be expanded to order 
$N-1$ \cite{Fry93}. In this paper we
will only study the clustering to order $N=3$ which means that bias only
needs to be quadratic in the local model. Thus, in practice, we will
be testing the following model:

\bea{eq:epsilon}
\delta_h&=&b_0+b_1\delta_m+\frac{b_2}{2}  \delta^2_m+\epsilon \nonumber \\
&=& b_1\delta_m+\frac{b_2}{2} (\delta^2_m-\sigma_m^2) +\delta_\epsilon
\eea
where $\epsilon$ represents the scatter around the local relation (and
also includes higher order contributions in $\delta_m$).
Because we require $<\delta_h>=0$, we have $b_0=-b_2 \sigma_m^2-<\epsilon>$
and define $\delta_\epsilon \equiv \epsilon-<\epsilon>$.

One important prediction of this local model is to expect the
shape of the correlation to be unaffected by bias (or in other words
that the effective bias is constant) on large scales:
\beq{eq:x2b}
\xi^h(r_{12}) \simeq ~b_1^2 ~\xi(r_{12}) + {\cal O}[\xi^2(r_{12}) ]
\eeq
The next to leading
contribution to $\xi^h$ above is
proportional to $\xi^2$ and consequently negligible at large $r_{12}$
where $\xi<1$. This will be tested below in section \S\ref{sec:2pt}.

It is in principle possible to use the shape of $Q_3^h$ in simulations (or observations) to
separate $b_1$ from $c_2 \equiv b_2/b_1$. This is done by a fit of 
 the halo (or galaxy) measurements of $Q_3^h(\alpha)$ in  Eq.\ref{Q3G}
using the corresponding dark matter predictions or measurements $Q_3(\alpha)$.
Changing $b_1$ will
produce a distortion of the U-shape of $Q_3$ (as a function of
$\alpha)$, while $c_2$ only produces a  constant shift. Thus, unless $Q_3$ is constant within the
errors or $b_1$ is very large, one could simultaneously measure $b_1$
and $c_2$ from $Q_3^h$ \cite{Frieman94,Fry94}.
This idea will be tested below in  section \S\ref{sec:3pt}.
One could also predict $b_1 $ from the ratio
of the halo to dark matter correlations: $b_1^2 = \xi_h/\xi_m$, but this
requires knowledge of the normalization of the dark matter
clustering amplitude  in $\xi_m$, which is often what we want
to fit from observations.
The fit to $Q_3$ can produce an estimate of the 
linear bias $b_1$ which is independent of the overall amplitude of
clustering $\xi_m$, because the $Q_3$ prediction is independent
of the normalization. This approach has already been implemented for the skewness $S_3$
(Gazta\~naga 1994, Gazta\~naga \& Frieman 1994), the bispectrum 
(Frieman \& Gazta\~naga 1994, Fry 1994, Feldman et al. 2001, Verde et al. 2002) 
or the angular 3-point function (Fry 1994, Frieman \& Gazta\~naga 1999,
Gazta\~naga \& Scoccimarro 2005, Gazta\~naga et al. 2005).

\subsection{Measurements in $\delta_h$ vs $\delta_m$ scatter plot}
\label{sec:scatter}

We are interested in exploring and determining the local bias parameters directly.
In order to do so we will compare the halo density contrast $\delta_h$ 
with the corresponding local matter density fluctuation $\delta_m$ at
the same cells. We will do this for all cells in the simulation.
This will give us an scatter plot of the relation $\delta_h=F[\delta_m]$ 
from which we can obtain $b_1$ and $c_2$ by means of a least mean square fit to the local bias parabola from the Taylor 
expansion of $\delta_h$ in Eq.\ref{eq:epsilon}.

Scatter plots of halos of more than n=50 particles are shown in Fig.\ref{scatterzRs}
for a selection of sizes of the cubical cell ($l_c=24,48,128$ Mpc/h),
which correspond 
to spherical top hat volumes of 
radius $R_s=14.9,29.8,79.4$ Mpc/h as labeled in the figures. Left and
right panels show results for $z=0$ and $z=0.5$ respectively. It is
apparent how the quadratic bias $c_2$ changes sign from convex
($c_2<0$) to concave ($c_2>0$) as the redshift increases.

One prominent feature in the plots is the 
discreteness of the $\delta_h$ values, i.e., that we see horizontal
lines in the figures. This obviously comes from the fact that we have an integer number of halos in each cell. The step
in the halo density fluctuations is consequently $\Delta \delta_h=1/\bar{n}$,  where $\bar{n}$ is the mean number of halos
in the cells. This is the value of the Poisson shot-noise, 
which will decrease when increasing the cell size or when lowering
the mass threshold of halos, for we will have a larger $\bar{n}$. The matter density field is also discrete, but
because the large number of matter particles per cell this effect is not visible in the plots.

\subsection{Smoothing scale}
The bias parameters obtained from the least mean square fit  
depend on the size of the cell used to smooth the density field, therefore the issue of 
what smoothing radius to use when comparing with clustering bias should be addressed. 
First, notice that for 
the smallest smoothing radius the scatter of points is very big. In this case many points have
$\delta_m \ge 1$, which situate us in a regime where the Taylor expansion of $F[\delta_m]$ can not be applied.
When the radius is set to a larger value the scatter gets reduced and almost all points have $\delta_m < 1$, 
situating us within the perturbation regime and producing a particular fit of the bias parameters. 

Our results of the dependence of the bias on 
the smoothing radius for several halo minimum masses are presented in
Fig.\ref{figb1Rs}.

As expected, we see that the values of $b_1$ and $c_2$ change significantly as we increase the smoothing radius
$R_s$ from $5$ to $20-25$ Mpc/h, from where they start to converge to their large scale values. The convergence
is reached faster at lower mass thresholds and, for a fixed mass, at lower redshifts.
In our study we will take smoothing radius of $30$ and $60$ Mpc/h, 
where the convergence regime has been reached. 

Through all the paper errors on the measured bias parameters have been computed using the jack-knife method with 64 subsamples
of the density fluctuations field. This is, we first compute the density fluctuations using the true mean density
of the simulation and then we create the jack-knife subsamples from which we obtain, using these fluctuations,
a set of 64 bias. Applying equation \ref{jkeq} gives the estimated jack-knife error. 
We have check that changing the number of regions does not change results significantly.


\subsection{Comparison with 2-point correlations}
\label{sec:2pt}

\begin{figure}
\center
\vskip -0.3cm
\includegraphics[width=77mm]{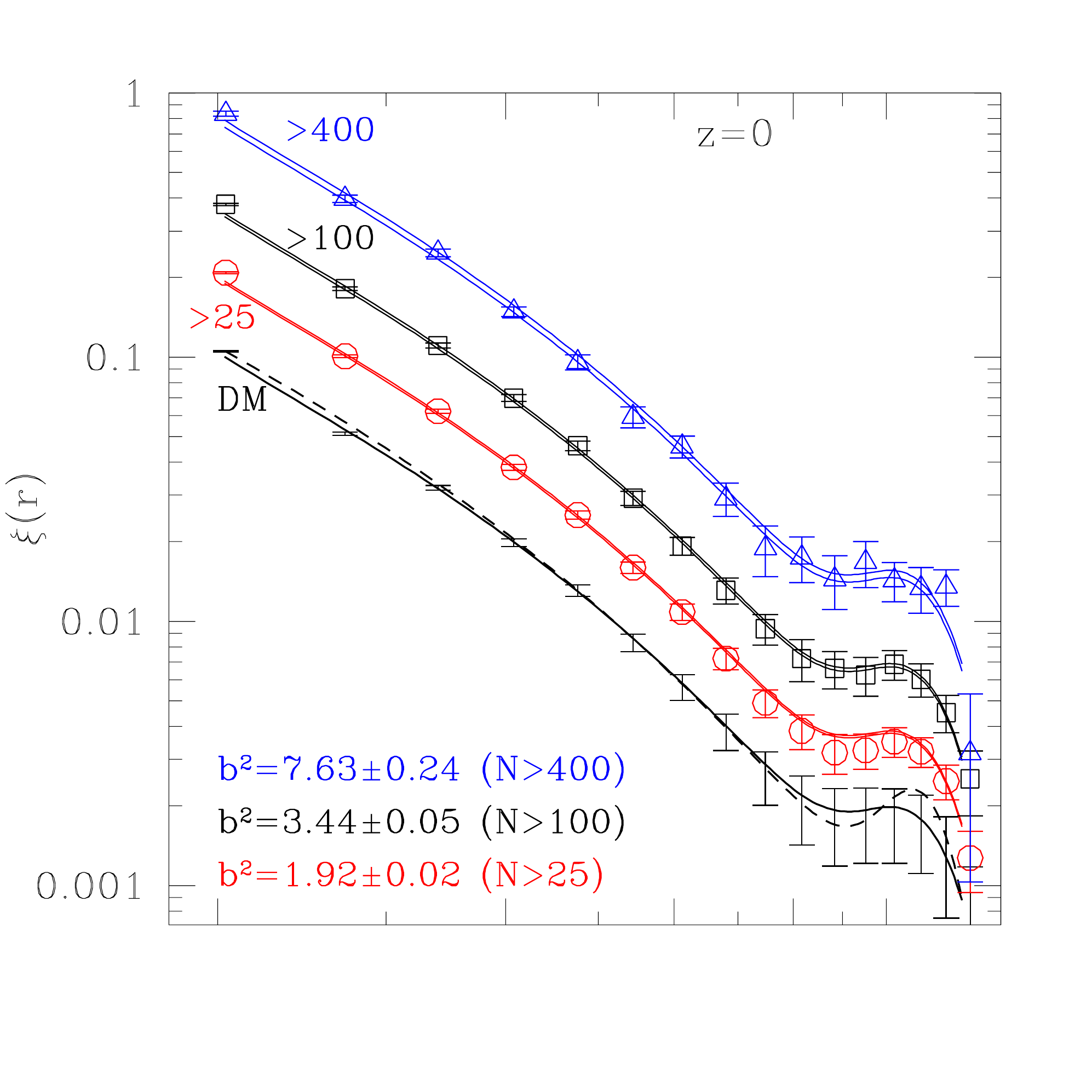}
\vskip -1.9cm
\includegraphics[width=77mm]{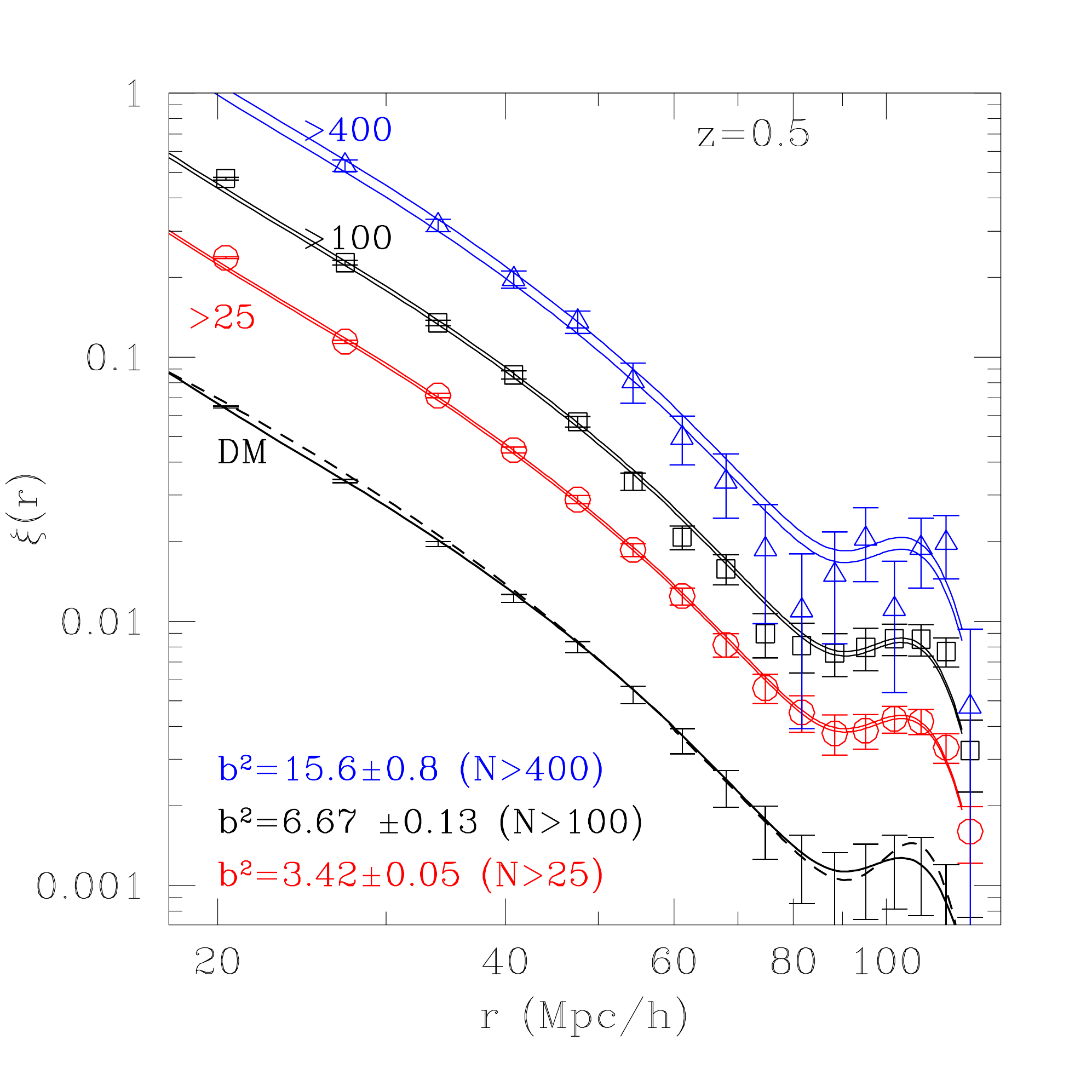}
\vskip -0.4cm
\caption[2-point correlation function]
{Symbols with JK errorbar show the 2-point correlation function $\xi(r)$ 
from simulations for different minimum number of particles ($N>25, 100$ or $400$) 
per halo as labeled in the figure. The bottom errorbars corresponds to the measurements in the
DM distribution. The bottom continuous (dashed) lines in each panel shows the RPT (linear)
theory prediction. The upper continuous lines show the best fit amplitude for the RPT
prediction shape, whose amplitudes $b^2$ are shown in the bottom labels. 
Top (bottom) panel correspond to z=0 (z=0.5).}
\label{correlacio3D}
\end{figure}

\begin{figure}
\center
\vskip -0.3cm
\includegraphics[width=77mm]{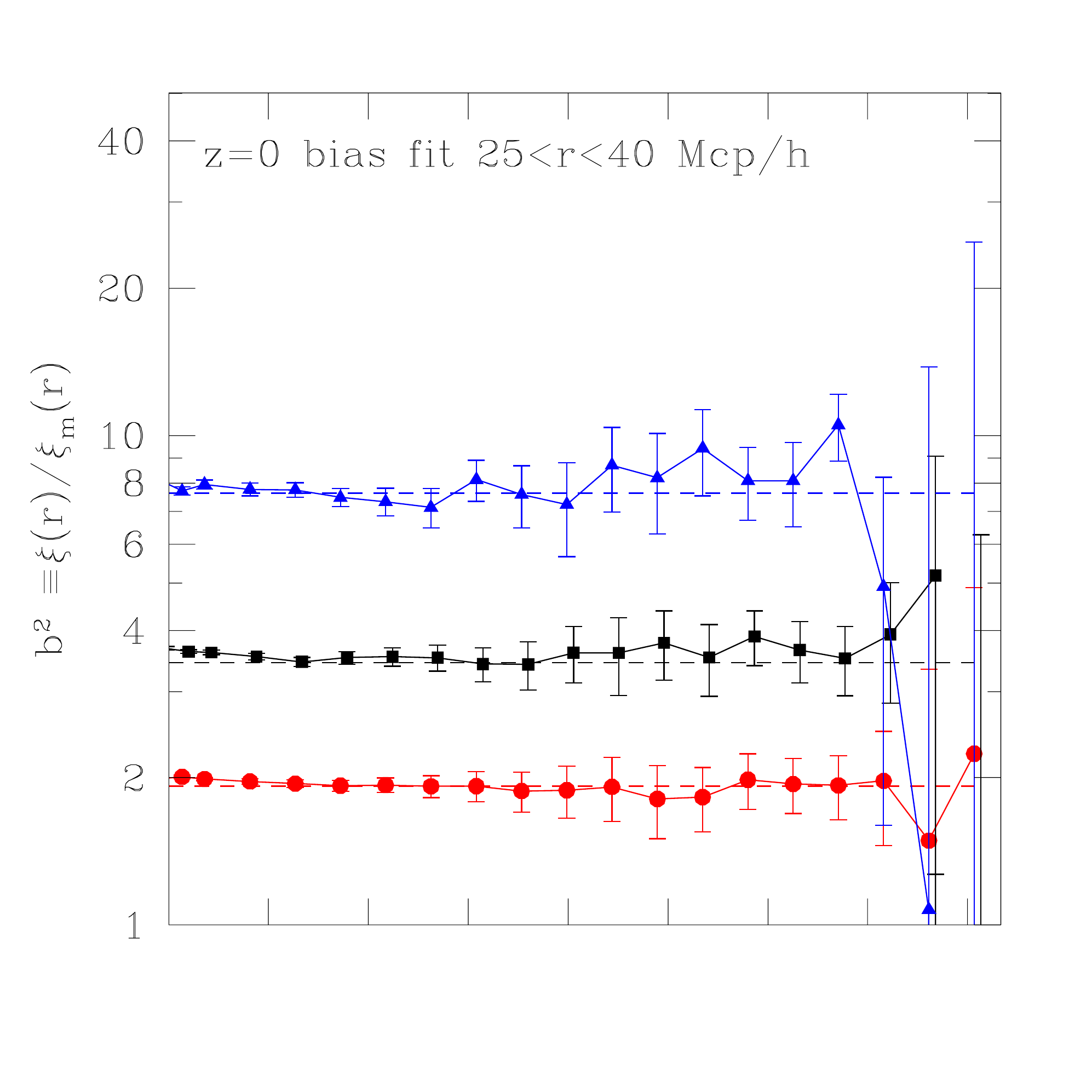}
\vskip -1.9cm
\includegraphics[width=77mm]{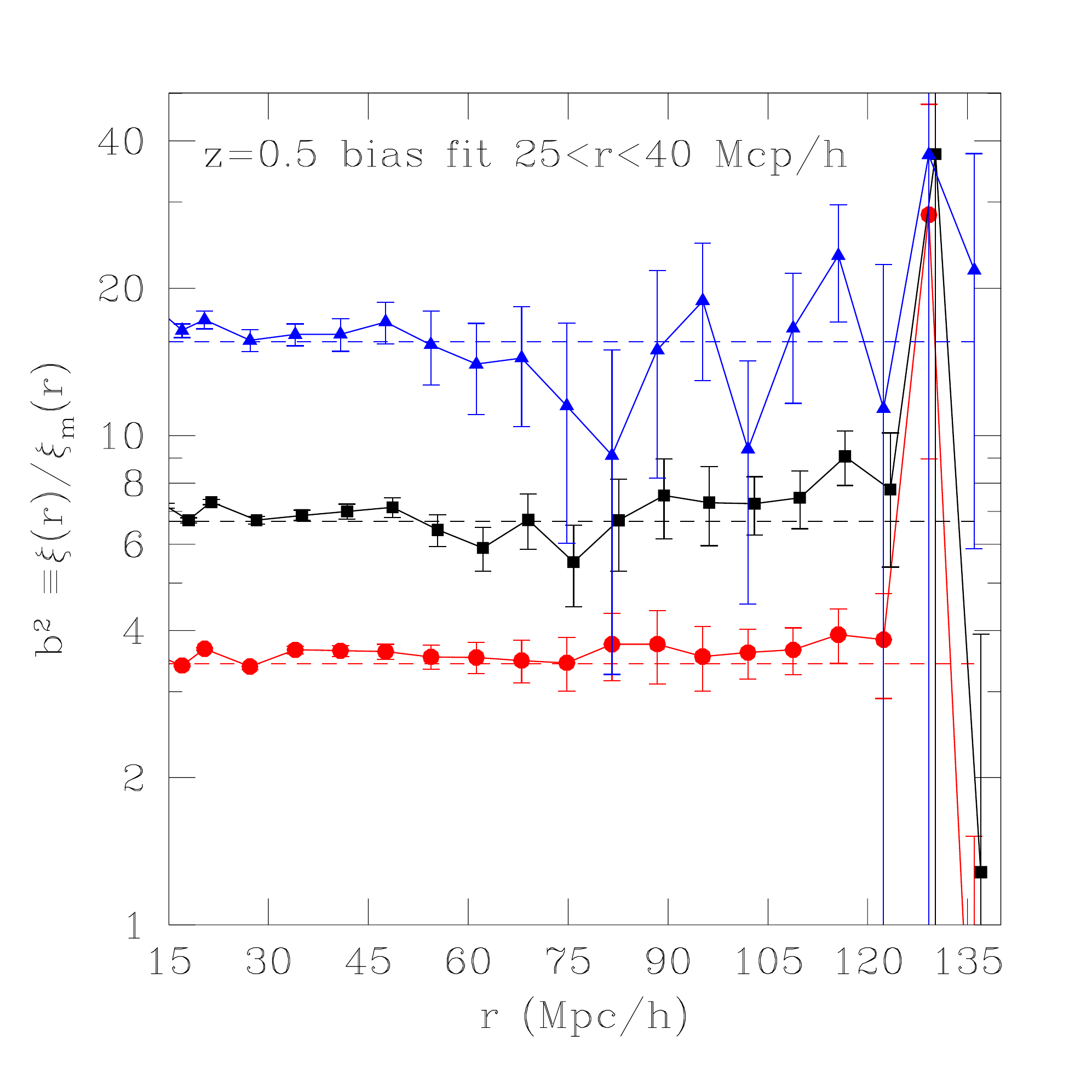}
\vskip -0.4cm
\caption[2-point correlation function for halos in the simulation]
{Bias from the ratio of 2-point correlation function $\xi(r)$ for different minimum number of particles
per halo N=25, 100, 400 (from bottom to top). 
Top panel shows results for $z=0$ and bottom panel for $z=0.5$. The dashed lines show the values
of the linear bias fit in the range $25<r<40$Mpc/h.}  
\label{biascorrelacio3D}
\end{figure}


We have computed the 2-point correlation function $\xi(r)$ for the matter
and halo density contrast in the simulation. To estimate $\xi(r)$ we
have used the 4Mpc/h density mesh of the simulation and average all the mesh
points separated by $(r\pm \Delta r)$, where $\Delta r = 0.5 Mpc/h$
(see Barriga \& Gaztanaga 2002 for details).
The results for the matter correlation function and for different halo masses
(given by the minimum number of particles per halo) are shown in 
Fig.\ref{correlacio3D}. The top panel shows the $z=0$ case and the bottom panel the $z=0.5$.
To convert particles to halo mass remember that $M_p = 2.34\;  10^{11} M_{\odot}/h$.  

As expected the more massive the halos the more biased the correlation function.
Note as well what is called the stable clustering, i.e., the fact that for a
given halo mass threshold the absolute value of $\xi$ remains approximately constant
in redshift while the matter correlation function decreases (in redshift). This could be
understood however because halos of a given mass  but at different redshift
do not correspond to the same Lagrangian mass. The ones at higher
redshift are situated in a rarer (less expected) matter fluctuations, being therefore more biased.

The measured correlation function from the simulation shows very clearly the acoustic peak at
about $\sim 110 Mpc/h$ for both the matter and the halo functions.
For comparison, in this figure we
have also plotted the Linear Perturbation Theory (PT) prediction
(dashed lines) and the Renormalized Perturbation
Theory (RPT) prediction (continuous) for the correlation function, which has been kindly provided by M. Crocce
The RPT shows deviations of the linear theory at much larger scales that have been previously
thought and even in the acoustic peak scale one gets a contribution of the nonlinear effects
\cite{Crocce08}. As can be seen in the figure these nonlinear contributions results into a smoother
prediction for the acoustic peak shape in the RPT that is in better
agreement with what we find in the simulations.

We find the bias from
\beq{bfromxi}
b(r)=\sqrt{\frac{\xi^h(r)}{\xi^m(r)}}
\eeq
This bias is expected to be constant at large scales in the local bias
model of Eq.\ref{eq:x2b}.
Cosmic variance and shot-noise will add variations to this large scale
constant bias.  Both errors get more pronounced
for larger scales (where we have few modes in the simulation) and for larger halo mass
thresholds (since the number of halos is smaller). This can be seen in 
Fig.\ref{biascorrelacio3D} where we plot  $b^2$ as a function of separation
for redshifts $z=0$ and $z=0.5$ and 
different mass thresholds. We do not find any evidence in the data for scale variations
of $b$ for $r>20$ Mpc/h. This favors the local bias model, but note that this statement is only 
accurate within the  $\simeq 10\%$ accuracy in our analysis. 

\begin{figure}
\center
\includegraphics[width=80mm]{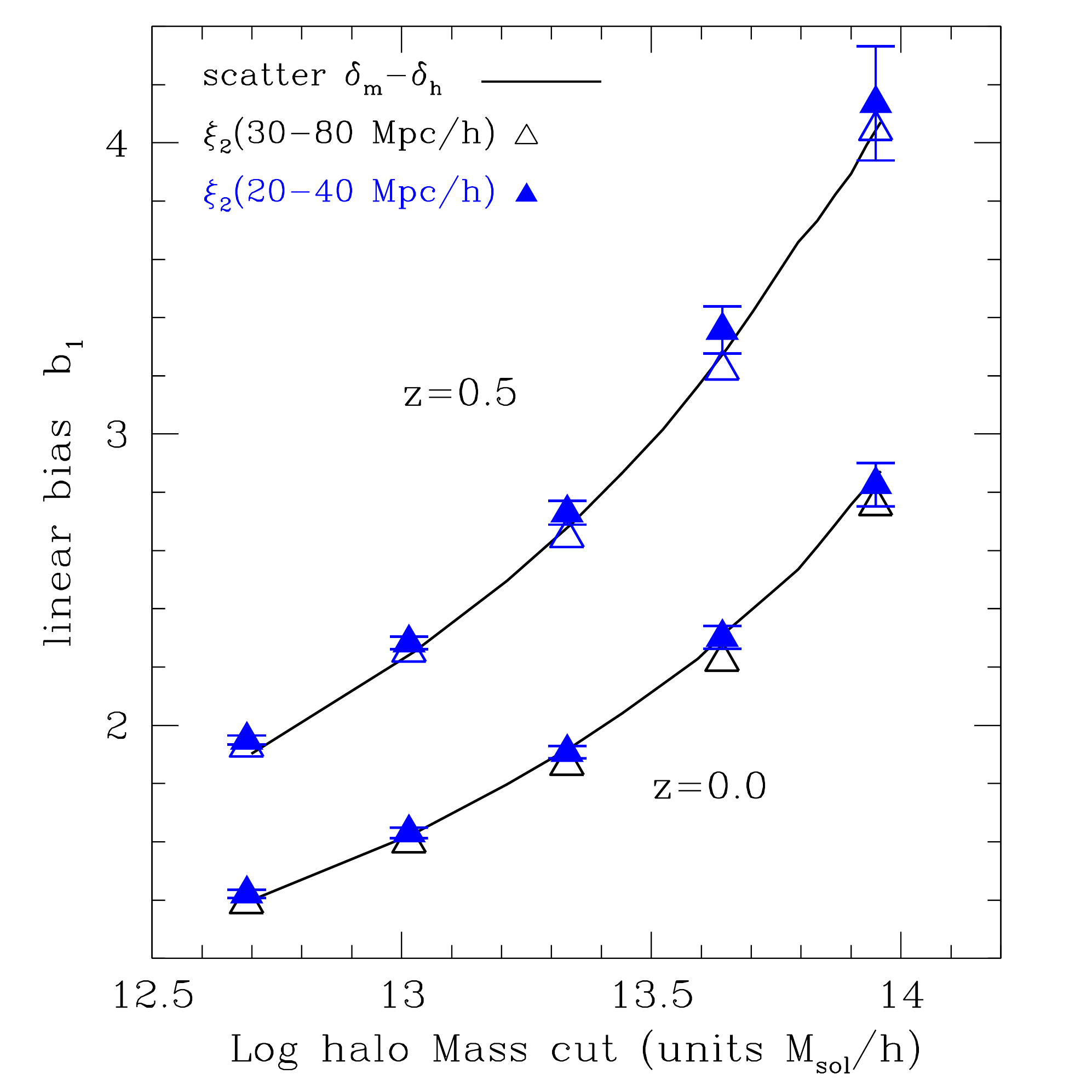}
\caption{Comparison of different estimates for the linear bias
as a function of the minimal halo mass.
Continuous line correspond to the local model fit to the
scatter relation $\delta_m-\delta_h$ in Fig.\ref{figb1Rs} at R=60Mpc/h.
Triangles correspond to bias from the 2-point function 
on large 30-80 Mpc/h scales  (open triangles)  and intermediate 20-40 Mpc/h
 scales (filled triangles).}
\label{fig:b1x2}
\end{figure}

\begin{figure*}
\center
\includegraphics[width=85mm]{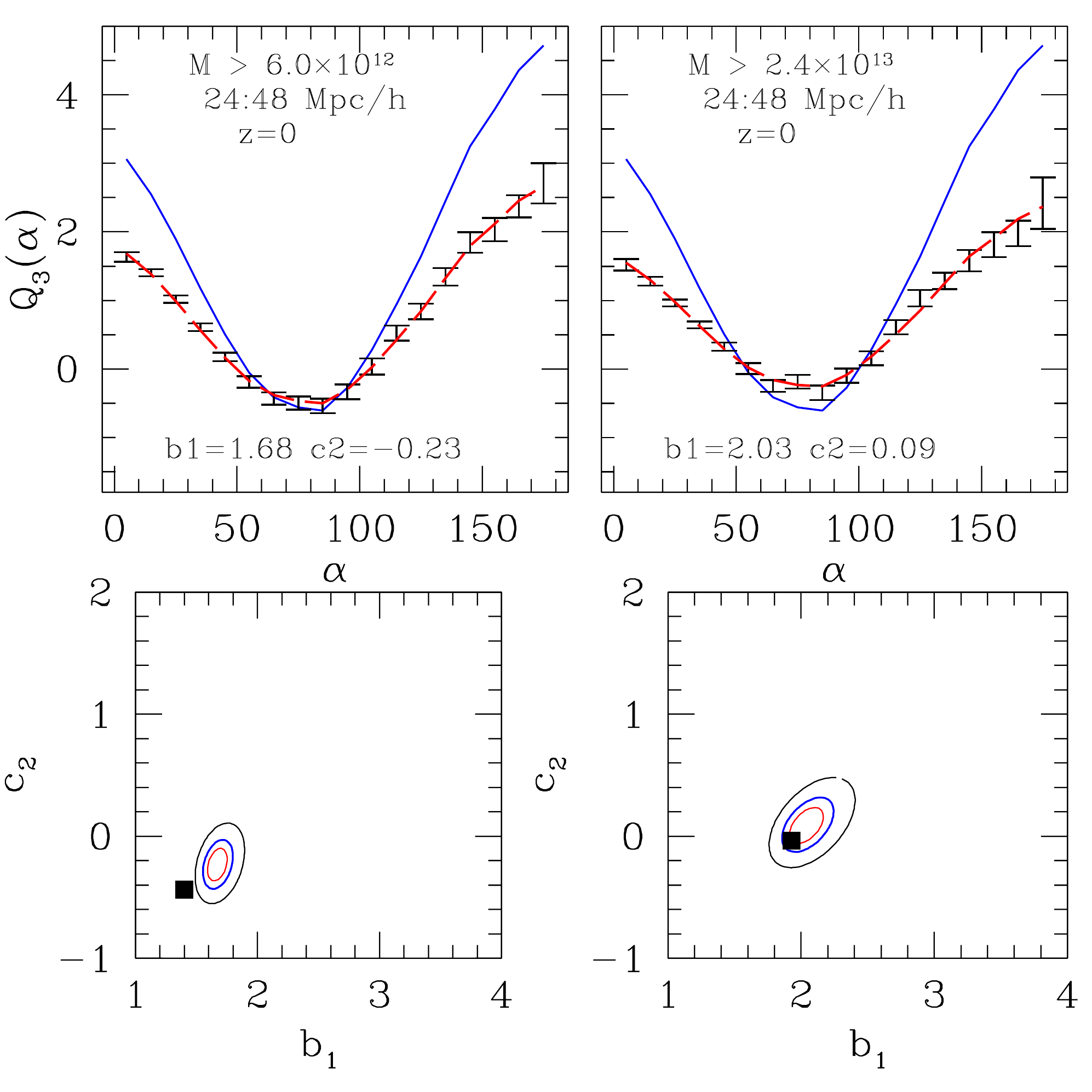}
\includegraphics[width=85mm]{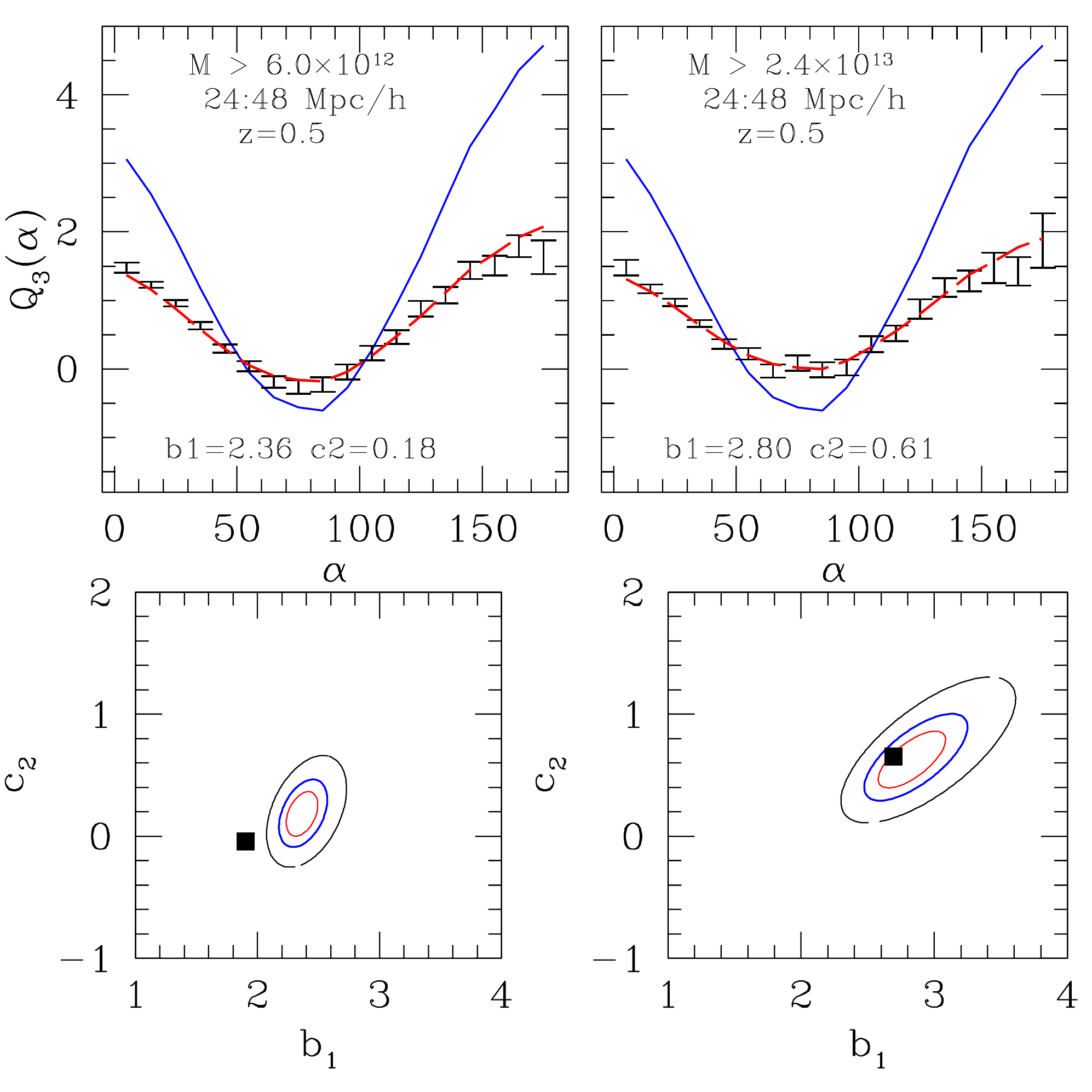}
\caption[Hierarchical relation  $Q_3$ from matter and halos at z=0]
{ Halo biases from the reduced 3-pt function $Q_3(r_{23},r_{12},\alpha)$ 
for z=0 (left set of panels)   and z=0.5 (right set of panels) with 
$r_{23} = 2 r_{12} = 48$ Mpc/h. Each column corresponds to
a different halo mass (as labeled in the top panels).
 {\it\bf TOP}: $Q_3^m$ in 
dark matter as measured in the simulation (blue lines)
 as compared to $Q_3^h$ in halos (errorbars) of the same
simulation. Long dashed red lines show 
the local bias model predictions (equation \ref{Q3G}) for the best fit
values of  $b_1$ and $c_2$ shown in the bottom panels. {\it\bf BOTTOM}:
$\chi^2$ contours in the $b_1-c_2$ plane for
$\Delta\chi^2=1, 2.3$ and $6.17$.
Best fit values are found by matching the measured $Q_3^h$ in
halos (symbols in top panels) with predictions in the local bias model, ie
$Q_3^h=(Q_3^m+c_2)/b_1$ where $Q_3^m$ are the dark matter
values ( blue continuous lines in top panels).
Filled squares show the values of $b_1-c_2$ from 
the local bias scatter plot $\delta_h-\delta_m$ in Fig.\ref{figb1Rs} at R=60 Mpc/h.}
\label{Q3z}
\end{figure*}

We do a fit to a constant $b(r)$, weighted by the inverse variance,
 for different range of scales. The result is
shown as continuous lines in   Fig.\ref{correlacio3D} and triangles in Fig.\ref{fig:b1x2}.
The bias from $\xi(r)$ is slightly larger when we fit to
smaller scales of $20-40 Mpc/h$, but results are consistent within errors.
We can see in this later figure that, within its errors, the bias
from clustering is in good agreement with the local bias determined
directly from the $\delta_h$-$\delta_m$ relation at larger scales.
The values in the figure correspond to cell size $R_s=60$ Mpc/h where the
bias in Fig.\ref{figb1Rs} has reach its asymptotic value for all
masses. The agreement
is no so good for smaller smoothing scales. Even for cells as large as $R_s=30$
Mpc/h we find some deviations in $b$ for large masses. 
This clearly indicates that the local bias prescription in
Eq.\ref{deltataylor} is to be understood as an effective relation
smoothed over very large scales and it fails when we try to apply it
as a truly local transformation (where $R_s<r_{12}$, with $R_s < 60$ Mpc/h).
Also note that at these large smoothings the stochastic component $\delta_\epsilon$ (see Eq \ref{eq:epsilon}) 
is small as illustrated in Fig. \ref{deltataylor} and that, in particular we can
neglect the stochasticity correlation between two different points 
$<\delta_\epsilon(r_1) \delta_\epsilon(r_2)>$ in the modeling of the
2-point correlation.


\subsection{Comparison with 3-point correlations}
\label{sec:3pt}

We have computed the hierarchical relation $Q_3(\alpha)$ 
(see equation \ref{Q3G} ) 
for dark matter and halos in the simulation
(as for 2-point function we follow Barriga \& Gaztanaga 2002).
We use triangles with fixed $r_{23}=2 r_{12} = 48 Mpc/h$
and $r_{13}$ given by the angle $\alpha$ between $r_{23}$ and $r_{12}$.
Some results for $z=0$  (left) and $z=0.5$ (right)  are shown in Fig.\ref{Q3z}. 
Dark matter measurements are shown as (blue) continuous lines
 while halo measurements correspond to  errorbars.
Errorbars in dark matter are negligible as compare to errors
in the halo distribution, which is dominated by shot-noise.
The standard perturbation theory prediction for $Q_3$ is quite
close to the DM measurements on these large scales.
Notice the characteristic U shape in $Q_3(\alpha)$. This is an indication of
filamentary structure, i.e, aligned structures
($\alpha \sim 0$, $\alpha\sim 180 \deg$ ) are more probable
than perpendicular configurations (for instance, equilateral triangles).
Spherical structures will produce constant values of $Q_3(\alpha)$.
As the bias increases, the distribution becomes less filamentary
and this information can be used to measure the bias.

We have fitted the shape of $Q_3^h$ in simulations to $b_1$ and $c_2$
in Eq.\ref{Q3G} using the corresponding dark matter measurements
$Q_3$ (we follow the procedure described in Gazta\~{n}aga \& Scoccimarro 2005)
Changing $b_1$ produces a distortion of the U-shape of $Q_3$,
while $c_2$ only produces a constant shift. The fits are shown as
contours in the bottom panel of Fig.\ref{Q3z} and they are compare
with the values of $b_1$ and $c_2$ (squares) from the scatter plot in
Fig.\ref{figb1Rs} at $R_s=60$ Mpc/h. For errorbars we use the JK
covariance matrix.  This matrix is degenerate because of the strong
correlations of different $\alpha$ bins. To be safe we only use the
two principal components with larger eigenvalues (see Gazta\~naga \&
Scoccimarro 2005). This is quite conservative in terms of the size of
the resulting errorbars. Better estimates would require a more careful
study of the covariance matrix, which is beyond the scope of this
paper.

\begin{figure}
\center
\includegraphics[width=80mm]{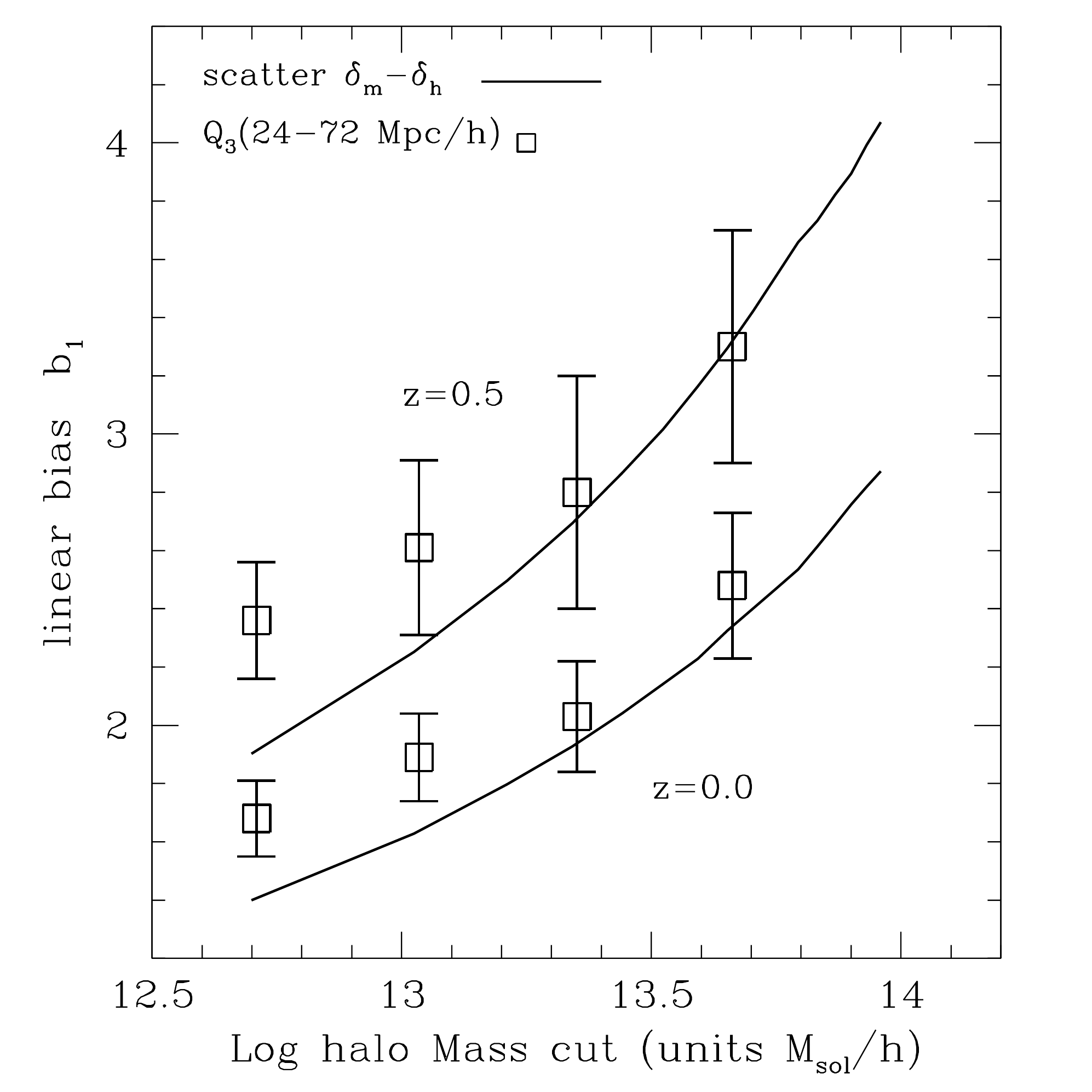}
\caption{Comparison of different estimates for the linear bias
as a function of the minimal halo mass.
Continuous line correspond to the local model fit to the
scatter relation $\delta_m-\delta_h$ in Fig.\ref{figb1Rs} at $R_s=60$Mpc/h.
Open squares come from fitting the 3-pt
 function $Q_3$, i.e., see Fig.\ref{Q3z}.}
\label{fig:b1q3}
\end{figure}

The values of $b_1$ recovered from $Q_3$ (squares) for different
mass thresholds are shown in Fig.\ref{fig:b1q3}. There is good
agreement in the general tendency of $b$ as a function of mass but
there are some significant deviations for small masses ($\log M<
13$). This failure of the local biasing model for $Q_3$ is intriguing
in the light of the very good agreement that we found from $\xi$ in
Fig.\ref{fig:b1x2}.  This is an important point to clarify because we
do not know $b_1$ in the real universe and we were hoping to be able
to use the values of $b_1$ from $Q_3$ to find the dark matter
normalization of $\xi$. According to Fig.\ref{fig:b1q3} this will
produce a significant (2-sigma level) deviation for small halo masses.

This mismatch can hardly be attributed to the stochastic component
$\delta_\epsilon$ (which includes also non-local contributions). As in
the case of the 2-point function, because the smoothing radius in the
local model is very large, we expect the stochastic correlation
components to be subdominant (see section above).  A key difference
between the 2 and the 3-point function is that the former takes
isotropic averages while the later keeps anisotropic information
(something which is not captured either by the skewness, see below,
which is a third order statistics but is smoothed in spherical cells).
So our finding hint in the direction that we need some anisotropic
component to the halo biasing model in Eq.\ref{deltataylor}, at least
for $\log M\simeq 13$. This conclusion might not be generic. For
biasing in galaxy mock catalogs where $b \simeq 1$, corresponding to
lower mass thresholds in the halo picture, Gazta\~naga \& Scoccimarro
(2005) and Marin et al. (2008) found good agreement of the values of
$b_1$  coming from $\xi$ and $Q_3$ clustering under the local model. 
More work needs to be done to clarify these issues.

\subsection{Comparison with the variance and skewness} 

So far we have studied
the bias from 2-pt and 3-pt correlation functions because they do not
suffer from  the discreteness effects that appear in  the variance and the skewness. 
 However, the latter are closer to the local model assumptions
(since they prove a local smoothed quantities). Since they bring different aspects to the 
comparison we will also study them here.   

We define the variance $\sigma^2$ and skewness $m^3$ as second and
third order moments of the fluctuation field:

\beq{varskew}
\sigma^2=\langle \delta^2 \rangle = \frac1N \sum_{i=1}^N \delta_i^2 ~;~
m_3=\langle \delta^3 \rangle = \frac1N \sum_{i=1}^N \delta_i^3 
\eeq
where the sum is over a fair sample of points in the simulation (ergodic assumption). 
In this case, one typically considers these quantities as a function of the smoothing radius R.
It is also convenient to define the normalized skewness:
\beq{skewnessnorm}
S_3  = \frac{m_3}{\sigma^4} 
\eeq

\subsubsection{Variance}
One of the common ways of determining the linear bias of galaxies or halos is by
comparing their variance with the measured/predicted matter variance.  However, to do
the correct comparison one has to account for the  shot-noise
contribution (a very similar problem occurs in the estimation of the power
spectrum which is the variance in Fourier space).
This contribution
to the variance appears because galaxies and halos are not a continuous fields, but discrete ones.
For a top hat window function,  $W_R(r)=\Theta(\mid r \mid -\, R )$ the Poisson shot-noise is well known and it is equal
to $1/\bar{n}$ where $\bar{n}$ is the mean number of halos in a sphere of radius R. The shot-noise corrected variance is therefore:
\beq{varsn}
\sigma^2(R)=<\delta^2>-\frac{1}{\bar{n}}
\eeq
where R stands for the window function smoothing scale. 
The dark matter in the simulation is also a discrete field and, as mentioned before, it will have its own shot-noise correction, 
which will obviously be much smaller than the halo one due to its higher number density. Now, we can use the halo variance
to compute an estimator for the linear bias as
\begin{equation}
b_{hh}\equiv\frac{\sigma(R)}{\sigma_m(R)}
=\sqrt{ \frac{<\delta_h \delta_h>-1/\bar{n}}{<\delta_m \delta_m >}}
\label{varianceb1}
\eeq
Another estimator for the linear bias that can be computed from the simulation is
\begin{equation}
b_{hm}\equiv\frac{<\delta_h \delta_m>}{<\delta_m \delta_m>}
\label{bilinear}
\end{equation}

In Fig.\ref{b1massclustering} we show the values for the different bias estimators
computed in cubical cells of side $l_c=48$ Mpc/h. 
The variance in a cubical cells is very similar to the one
in a top hat smoothing sphere of equal volume 
as the cube (Baugh, Gaztanaga \& Efstathiou 1995). 
For our cells of side $l_c=48$ Mpc/h the spherical equivalent
radius is $R=29.8 Mpc/h$. Errors in the figure are from the 
Jack-knife method with 64 regions, and we have checked that 
changing the number of regions does not change results significantly.  

\begin{figure}
\vskip -.5cm
\hskip -.5cm
\includegraphics[width=100mm]{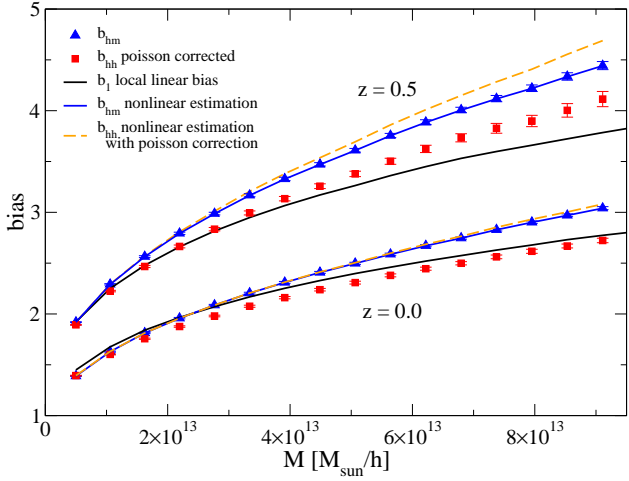}
\caption[Comparison of $b_1$]{
Bias as a function of halo mass.
The linear bias $b_1$  (shown as black lines) 
is estimated from a fit to the scatter plot $\delta_h$-$\delta_m$
in  the simulations. 
This is compared with the bias values obtained from the
(shot-noise corrected) 
variance $b_{hh}=\sigma/\sigma_m$ (red squares) and
from the cross-correlation $b_{hm}=<\delta_m \delta_h>/\sigma_m^2$ 
 (blue triangles). 
Also shown are the predictions for $b_{hh}$ and $b_{hm}$ after applying
non-linear contributions (i.e.,  in Eq.\ref{blpred} and Eq.\ref{bcpred}).
Results are shown for both $z=0$ (bottom
lines and symbols) and $z=0.5$ (on top).}
\label{b1massclustering}
\end{figure}

We can see that all three bias estimators $b_1$, $b_{hh}$ and $b_{hm}$
give significant different results given the errorbars. Consequently one needs 
to be cautious when trying to use these bias estimators for precision cosmology where
errors lower than 10\% are sought. Below we discuss the origin of these differences
focusing mainly in non-linear and discreteness effects, which we find
are the dominant effects.  Other contribution could arise from the
 truncation of the Taylor expansion.

\subsubsection{Skewness}

\begin{figure}
\hskip -0.8cm
\includegraphics[width=100mm]{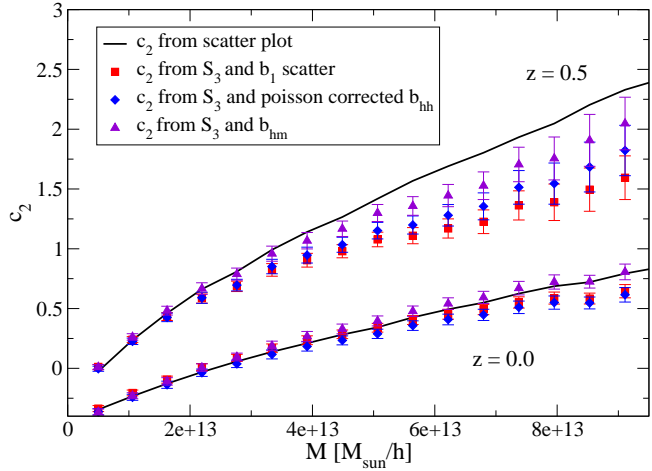}
\vskip -0.4cm
\caption[Comparison of $b_1$ ]
{ Dependence of $c_2$ on the halos mass as measured directly from the $\delta_h$-$\delta_m$
local relation in the simulation (black lines) compared with the values obtained from the
skewness and three different linear bias estimates in Eq.\ref{skewc2}:
$b_1$ from the local relation (red squares),  $b_{hh}$ 
 (blue diamonds) and $b_{hm}$ (pink triangles).
 Top (bottom) set of lines are for $z=0$ ($z=0.5$).}
\label{c2massclustering}
\end{figure}

An important
clustering statistic for understanding quadratic bias is the skewness. 
As all the moments and cumulants of the halo field it has to be shot-noise corrected. For the normalized
skewness this correction is found to be (eg. see Gaztanaga 1994):
\beq{S3shonoise}
S_3(R)=\frac{<\delta^3>-3\sigma^2(R)/\bar{n}-1/\bar{n}^2}{\sigma^2(R)}
\eeq
where $\sigma^2$ is again the shot-noise corrected variance and $R$ stands for the window function 
smoothing scale.
Note that when comparing the measured skewness to predictions one has to take into account the fact that we are smoothing
the density field. For a top hat smoothing and  CDM power spectrum
the normalized skewness can be approximated by \cite{jusz93,CooraySheth02,bcgs}
\beq{skewscale}
S_3=4+\frac{6}{7}\Omega_m^{-2/63}+\gamma_1
\eeq
where $\gamma_1=\frac{d ln(\sigma^2 (R))}{d ln(R))}$.
Obviously for the Einstein-de-Sitter cosmology and no smoothing we
recover the well known value in the spherical collapse model
$34/7$ (see Fosalba \& Gaztanaga 1998 for the interpretation
in terms of the spherical collapse model).

With the skewness and the linear bias we can easily compute $c_2$ as
(see section \S2)
\beq{skewc2}
c_2=(S_3^h \,b_1-S_3^m)/3
\eeq
Here we can either use the direct local $b_1$ as measured from the $\delta_h-\delta_m$ scatter plot 
or other estimators of the linear bias as $b_{hm}$ or $b_{hh}$. Results are shown in Fig.\ref{c2massclustering}
and compared with the $c_2$ obtained directly from the $\delta_m - \delta_h $ scatter plot fit. 
Errors for these points are computed by means of the Jack-knife method with 64 subsamples in the simulation.
As in the case of the variance we find significant deviations between
the different estimators. Next-to-leading order contributions as well as modeling stochasticity
would be needed for precision cosmology.

\subsection{Non-Linear effects and stochasticity}

In order to asses how good the linear approximation is we compute
the nonlinear contribution to the linear bias $b_{hm}$ and $b_{hh}$
using Eq.\ref{eq:epsilon}. We start with $b_{hm}$ which should be
subject to smaller discreteness effects. The next order in 
$\sigma^2$ is:

\begin{equation}
b_{hm} = \frac{<\delta_m \delta_h>}{<\delta_m \delta_m >} = b_1 + \frac{1}{2} b_2 S_3 \sigma_m^2
+ b_{\epsilon}
\label{blpred}
\end{equation}
where $b_{\epsilon} \equiv {{<\delta_m\delta_{\epsilon}>}\over{<\delta_m\delta_m>}}$, 
and $\delta_{\epsilon}=\epsilon-<\epsilon>$. Because of symmetry reasons, $b_\epsilon$
can be expected to be very small, as we will show next.  

Nonlinear corrections in Eq.\ref{blpred} seem to account  well for the difference 
that we saw in Fig.\ref{b1massclustering} 
between the measured $b_{hm}$ (blue triangles) and the linear bias
$b_1$ (black continuous line) 
from the fit to the scatter plots.  
The non-linear correction to $b_{hm}$ in Eq.\ref{blpred} is also shown in  Fig.\ref{b1massclustering} 
as a blue line (for $b_\epsilon=0$) and it overlaps well with the
$b_{hm}$ measurements  within errors. 
The nonlinear terms are therefore large  (10-15\% effect) 
 and certainly have to be taken
into account in precision cosmology. 
We infer from this very good agreement that the contributions from the scatter
$b_{\epsilon}$ in Eq.\ref{blpred} and the effect of the Taylor
truncation (i.e.,  higher orders in the expansion) are negligible given
the errors.

The corresponding corrections for $b_{hh}$ is:
\begin{equation}
b_{hh}^2 = b_1^2 +  \left[b_1  S_3 +
          \frac{1}{2} b_2
            + \frac{1}{4} b_2 S_4 \sigma_m^2\right] b_2\sigma_m^2  + \mathcal{E}_{hh} \nonumber \\ 
\label{bcpred}
\end{equation}

where the second term includes all the non-linear corrections and the
third term is:
\begin{equation}
\mathcal{E}_{hh}=b_1 b_{\epsilon}+{<\delta^2_m
\delta_{\epsilon}>\over{\sigma_m^2}}+{<\delta_{\epsilon}^2>-1/\bar{n}\over{\sigma_m^2}}
\label{ehh}
\end{equation}
which only includes terms involving the scatter. As pointed out above, because of
symmetry, we expect linear terms in $ \delta_{\epsilon} $ to vanish so that
$< \delta_{\epsilon} \delta_m^n > \simeq 0, n=0,1,2$. This is well supported
by the good agreement that we found above between $b_{hm}$ 
in Eq.\ref{blpred} (with $b_\epsilon \simeq 0$)
and measurements in Fig.\ref{b1massclustering}. But
this might not be necessarily the case for the quadratic
term $<\delta_{\epsilon}^2>$ because there is no cancellation
between positive and negative fluctuations.
 The  $1/\bar{n}$ term comes from the shot-noise
correction  (i.e.,  Eq.\ref{varianceb1}) which
 allows us to move from the discrete to the continuous halo variance;
it assumes that halos are a Poisson sample of the dark matter field. 
If all scatter $<\delta_{\epsilon}^2>$ in the local relation were
just Poisson, then we expect that 
$\mathcal{E}_{hh} \simeq <\delta_{\epsilon}^2> -1/\bar{n} \simeq 0$.

In Fig. \ref{b1massclustering} we
show how the non-linear corrections in Eq.\ref{bcpred}
fail to explain the difference between $b_{hh}$ and $b_1$.
The predicted $b_{hh}$ (dashed line) is higher that $b_{hm}$
(blue triangles) while the measured one (red squares) is lower.
In fact, the nonlinear terms seem to increase
the differences between the predicted and measured bias.
This could be explained if $\mathcal{E}_{hh}$ turns out to be negative,
which would happen if the scatter is sub-Poisson (smaller than
Poisson) and consequently we overcorrected shot-noise  
it by using $1/\bar{n}$ term. Sub-Poisson shot-noise have been found 
in simulations (Casas-Miranda et al. 2002) for halos larger than
$M_\star$.\footnote{Also note that the same effect seems to result in
super-Poisson errorbars (Cabre \& Gaztanaga 2009). This two statements 
are not in contradiction because the former refers to the Poisson
correction to the mean variance (in Fourier or configuration space) while
the later refers to the noise or error (around the mean 2-point
function)  induced by discreteness noise.}

\begin{figure}
\includegraphics[width=90mm]{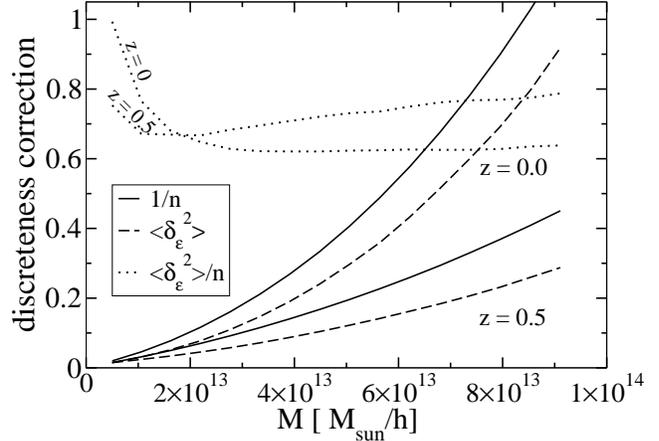}
\caption[Comparison of $b_1$ ]
{Comparison of the Poisson shot-noise correction $1/\bar{n}$ (continuous line)
and the scatter $<\delta_{\epsilon}^2>$ (dashed line) in the
local bias. There are two sets of lines, one for each redshift as
labeled in the figure (larger values correspond to $z=0$). 
Dotted lines show $\sigma_\epsilon^2$, the ratio of the two in
Eq.\ref{eq:shot-noise}.}  
\label{scattermoment}
\end{figure}

We have indeed found that our halo simulations have 
 $<\delta_{\epsilon}^2>$ which is smaller than $1/\bar{n} $. This can be seen in
Fig.\ref{scattermoment}, which compares the two terms. 
Besides shot-noise or discreteness effects $<\delta_{\epsilon}^2>$
also include other sources of scatter: 
non-deterministic bias and
possibly higher order contributions than the quadratic
terms in Eq.\ref{eq:epsilon}. 
 The later is a smoothed component
and is unlikely to result in a major increase in the actual scatter.
Fig.\ref{scattermoment} therefore indicates that the final scatter
is overestimated by the Poisson model.
We can write the new effective shot-noise term as:
\beq{eq:shot-noise}
<\delta_{\epsilon}^2> \equiv {\sigma_\epsilon^2\over{\bar{n}}}
\eeq
where $\sigma_\epsilon^2$ is plotted as dotted lines in Fig.\ref{scattermoment} .
 For small halo masses $\sigma_\epsilon^2$
tends to unity, while it is roughly constant
$\sigma_\epsilon^2 \simeq 0.6-0.8$  for larger masses.

    \begin{figure}
\vskip -0.5cm
\hskip -0.5cm
\includegraphics[width=100mm]{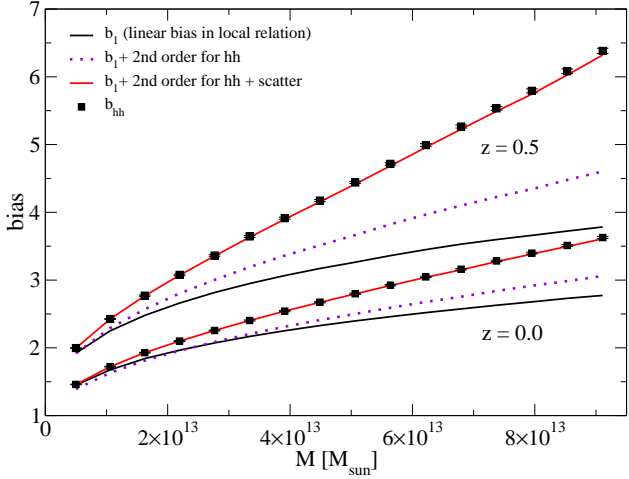}
\caption[Comparison of $b_1$]{
Bias in $b_{hh}$
as a function of halo mass as in Fig.\ref{b1massclustering}
but here we do not apply  the Poisson shot-noise correction to 
the measurements of $b_{hh}$.
When apply instead the discreteness correction to the predictions
(this correction estimated from the scatter $<\delta_{\epsilon}^2>$
shown in Fig.\ref{scattermoment}).
We only find a good agreement between the non-linear predictions (red lines) and
measurements (squares) after both discreteness and non-linear terms are
included.}
\label{b1massclustering2}
\end{figure}

In Fig.\ref{b1massclustering2} we apply the discreteness correction to the
prediction rather than to the measurements (which are not corrected
here for Poisson shot-noise).
When we use the  new estimate for the scatter,
i.e., $\mathcal{E}_{hh} \simeq <\delta_{\epsilon}^2>$ we find a very
good match between  the predictions and the measurements for $b_{hh}$,
We can see here that, as happened for $b_{hm}$ in Fig.\ref{b1massclustering}, 
non-linearities are also important for the variance. 
The main difference between  $b_{hm}$ and  $b_{hh}$
is that the later also needs a shot-noise correction 
that is different from Poisson, at least for large halo masses.
Both discreteness and  and non-linearities are needed to 
interpret the bias from the variance.

\subsection{Cross correlation and stochasticity}

A simple measure to study deviations away from the local linear bias
relation has been pointed out by several authors ( see Tegmark \&
Peebles 1998, Dekel \& Lahav 1999,  Seljak, Hamaus \& Desjacques 2009,
Cai, Bernstein \& Sheth 2010, and references therein).
This is to consider the dimensionless cross-correlation coefficient between
the distribution of mass and galaxies (we use halos as a proxy for
galaxies in our case):

\beq{stovar}
r \equiv\frac{<\delta_m\delta_h>}{\sqrt{<\delta_m\delta_m> <\delta_h\delta_h>}}
\eeq
which is in general a function of scale. 
In the local linear bias model $r=1$. But both non-linearities and
stochasticity (the scatter around the local relation) can change this
away from one. Note that this test is fundamentally different from
previous test of the local bias. This test focus on how important is
the stochasticity in the bias relation. For a deterministic function
one expects $r=1$, but when the scatter in the $\delta_h-\delta_m$
relation is large one would expect that it could have a different
impact in both parts of this ratio.

\begin{figure}[top]
\center
\vskip -1.4cm
\includegraphics[width=90mm]{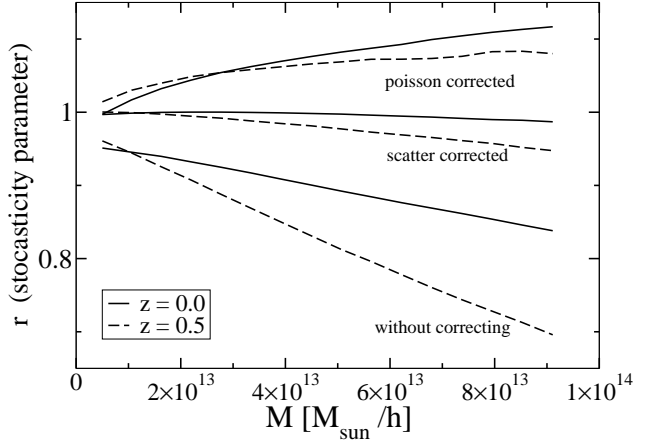}
\caption{Dimensionless cross-correlation coefficient $r$ in
 Eq.\ref{stovar} as a function of halo mass. This is for 1-point
fluctuations smoothed over cells of $R\simeq 30Mpc/h$ radius.
Different pairs of lines show results using different ways to correct
for the discreteness in the halo variance. Dashed (continuous) lines
correspond to $z=0.5$ ($z=0.0$).}
\label{stochasvariance}
\end{figure}

For 1-point smoothed fields
we have that in our notation (see Eq.\ref{varianceb1}-\ref{bilinear}) this corresponds to:

\beq{stovar2}
r=\frac{b_{mh}}{b_{hh}}
\eeq
We can estimate this quantity directly from the variance measured in 
simulations. The result is shown in  Fig. \ref{stochasvariance} as a function
of halo mass. We compare measurements without any correction
(lower lines) and using two different ways to 
correct for scatter and discreteness effects in 
the halo variance: the Poisson corrected variance:
$<\delta_h^2>-1/\bar{n}$ and the scatter corrected 
variance $<\delta_h^2> - <\delta_\epsilon^2>$.
The cross-correlation deviates significantly
from unity if we do not correct from these  effects. Deviations
increases with halo mass and redshift, and can be as large as 20-30\% 
for large halos.  As shown
before (eg. see Fig.\ref{scattermoment}) the Poisson model does not
provide a good correction for the scatter. If we use instead the
scatter away from the local relation, as measured in the simulation, we
recover values which are close to unity.

\begin{figure}
\center
\includegraphics[width=79mm]{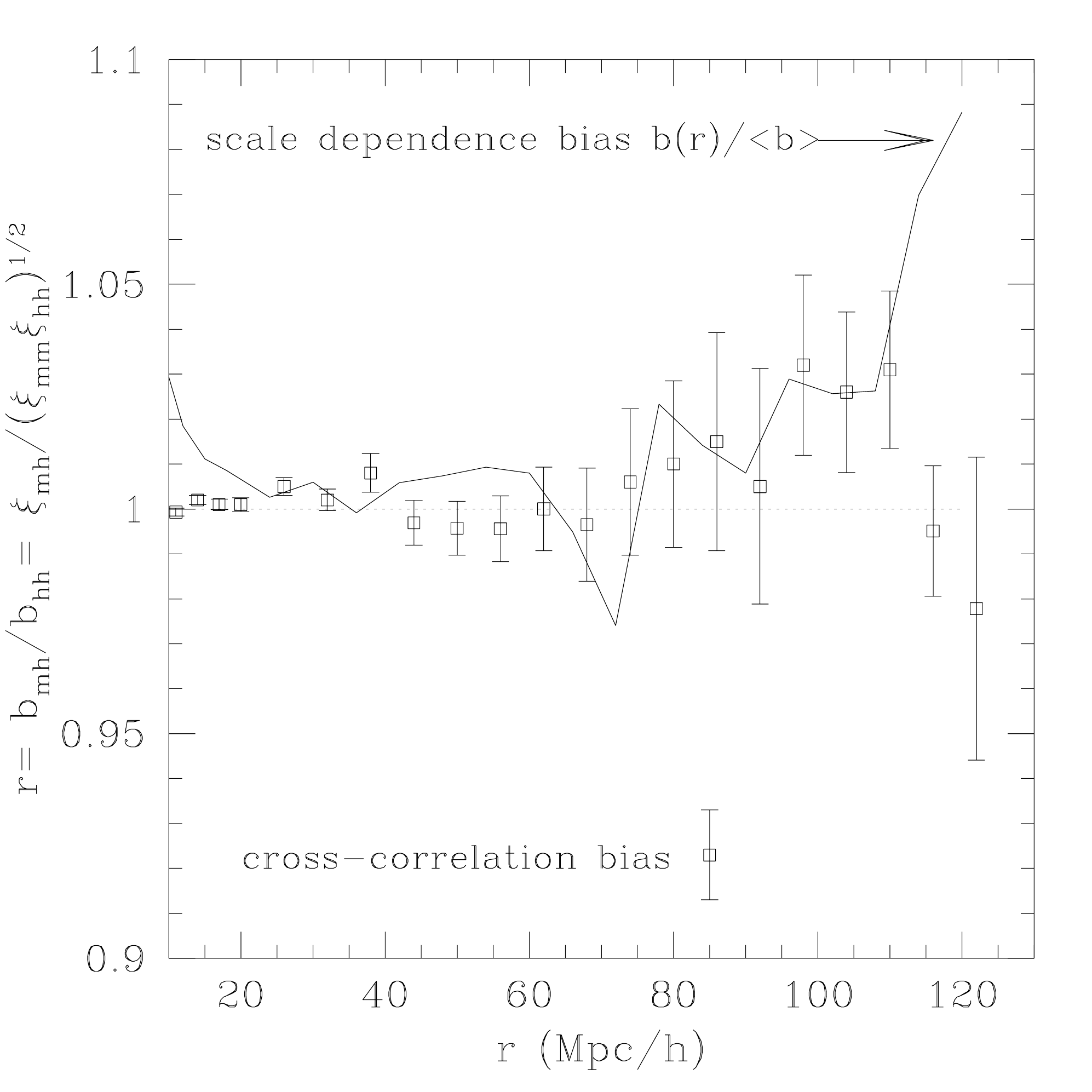}
\caption{Symbols with errorbars show $r$ in Eq.\ref{stovar}, i.e.,  the 
dimensionless cross-correlation coefficient  between  dark 
matter and halos with $M>5 \times 10^{12}$ in the MICE simulation at
$z=0.5$.
The continuous line correspond to the scale dependence bias $b_{hh}$,
normalized to the mean value. } 
\label{stochas2pf}
\end{figure}

We can also estimate $r$ in the 2-point correlation function, which should be
less affected by discreteness effects. 
Fig.\ref{stochas2pf} shows $r$ as a function of scale (separation
between pairs) for 
halos with $M>5 \times 10^{12} M_{\sun}$. 
In Fig.\ref{stochas2pf} we estimate JK errors from the $r$ ratio, i.e., we estimate the ratio in
different JK subsamples and calculate the error from the scatter in the
JK regions (this produces smaller errorbars because sampling variance
mostly cancels in doing the ratio). For comparison we also show
in this figure (continuous line) how much $b_{hh}$ deviates from 
a constant (i.e., from Fig.\ref{biascorrelacio3D}).
The measurements are compatible with
unity for all scales. There is a hint of a deviation ($\simeq 3\%$) around the BAO
scale which could be related to recent findings about scale
dependence bias (eg. see Desjacques et al. 2010 and references therein).

Similar results, but with much larger errors, are found for larger halos masses and different redshifts. 
Note that for masses larger than $10^{14} M_{\sun}$ 
Manera et al. 2010 found $b_{hm}/b_{hh}$ to be slightly larger
than unity, with $b_{hm}$ measured at low k in Fourier space and
$b_{hh}$  at large separations in the autocorrelation function.

Altogether, our analysis indicates that the linear local bias model provides
a very good approximation, within our sampling errors,  for the 2-point
function. On scales larger than $r \simeq 20 Mpc/h$,
the halo-halo correlation and halo-mass correlations
are, to a good approximation, linear tracers of the underlying
dark-matter correlation function and the resulting bias is
just the one expected in linear theory. This conclusion
is important to interpret measurements of redshift space distortions
and BAO  in galaxy surveys, which on large scale usually are
interpreted under the  assumption that linear bias and linear theory
are good approximations.

This is not so much the case for the variance, which seems more
affected by non-linearities and discreetness effects. This is
understood from the fact that the variance (as well as the power
spectrum) is quadratic in fluctuations and is an average over all
scales, including small, non-linear, scales.


\section{Bias Predictions form the Mass function and their performance}

In the previous section we have studied how the local bias performs 
when compared with the bias measured from clustering. In this section
we will compare those bias with predictions from the mass function. 

\subsection{Bias predictions from the mass function }

In the peak-background split Ansatz (Bardeen et al. 1986; Cole \&
Kaiser 1989) one can relate the halo bias with the halo mass function at large scales
by treating perturbed regions as if they were unperturbed regions in a slightly 
different background cosmology universe but one of the same age (Martino \& Sheth 2009).

Consequently, from a well motivated functional form of the mass function,  one can
derive theoretical predictions for the halo bias parameters as well as study their accuracy
(Mo et al. 1997, Scoccimarro et al. 2001, Cooray \& Sheth 2002, Manera et al. 2010).

In this paper we will use the Sheth and Tormen (1999) mass function:
\begin{eqnarray}
n(m) dm & = &  {\rho_m \over m }f(\nu) d\nu \\
\nu f(\nu) & = & A(p) \left(1+(q\nu)^{-p}\right) \left(\frac{q\nu}{2\pi}\right)^{1/2}
\exp \left(-\frac{q\nu}{2}\right).
\label{eq:ST}
\end{eqnarray}
where
$ A(p)=[1+(2^{-p}\Gamma(1/2-p))/{\sqrt{\pi}}]^{-1}$. 
is the normalized amplitude. The corresponding bias predictions are
\bea{biasparam}
\label{eqbiasparam}
b_1(m,z) & = & 1+\epsilon_1+E_1 \nonumber  \\
b_2(m,z) & = & 2(1+a_2)(\epsilon_1+E_1)+\epsilon_2+E_2  
\eea
where $a_2=-17/21$, and 
\bea{biasparam2}
\epsilon_1 & = &\frac{q\nu-1}{\delta_{sc}(z)} \quad\quad
\epsilon_2=\frac{q\nu}{\delta_{sc}(z)}\left(\frac{(q\nu)^2-6q\nu+3}{\delta_{sc}(z)}\right)
\nonumber \\
E_1 & = & \frac{2p/\delta_{sc}(z)}{1+(q\nu)^p} \quad\quad
\frac{E_2}{E_1}=\frac{1+2p}{\delta_{sc}(z)}+2\epsilon_1 \\ \nonumber
\eea
Throughout the paper $\nu=\delta^2_{sc}(z)/(D^2(z)\sigma_0^2(m))$. In
this notation $D(z)$  is the growth factor in units of its value at 
$z=0$; $\sigma_0(m)$ is the linear variance of the matter field at redshift $z=0$, when  
smoothed with a top hat filter of radius
$R=(3m\bar{\rho}/4\pi)^{-1/3}$ 
and $\delta_{sc}(z)$ is the critical density contrast for collapse at a given redshift z. 
Although it is popular in the literature to use a fixed value for $\delta_{sc}$ we
will be using its proper redshift dependence from the spherical collapse (Eke et al. 1996, Cooray \& Sheth 2002) 
since there is some indications that in this case the mass function
closer to universal (Manera et al. 2010).  

These predictions for the bias depend on the mass function through the parameters $p$ and $q$. 
When $p=0$ and $q=1$ we recover the Press-Schechter (Press \& Shechter 1974) formula.
Original values for this mass function fit were $p\simeq0.3$ and $q\simeq0.7$ (Sheth \& Tormen 1999) 
which discusses afterwards that $q=0.75$ (and therefore $A\simeq0.3222$) gives better results
\cite{ST01,CooraySheth02}. We confirmed that this is the case and consequently we will use
the latter values as their fiducial values for the ST. At the same time
we will also use our own set of $p$ and $q$ values obtained by fitting the mass function
as we explain in section \ref{massfuncsection}. Bias parameters from a mass function with a functional
form like that Warren et al. (2006) has been studied by Manera et al. (2010) and showed to give
similar results than that of ST for a range of masses similar to that of this paper.  

The above $b_i(m)$ predictions are for a given halo mass, but the simulation results
are for halos above a mass threshold, consequently in this paper 
we integrated these predictions over the mass range, weighting appropriately according 
the number of halos at each mass.
If one has a model for populating galaxies in halos one can weight each halo 
by the number of galaxies given by the halo occupation distribution (HOD), and 
therefore obtain a prediction for the galaxy bias. 
For an approach of how this can be done see for instance, Sefusatti \& Scoccimarro (2005)
or Tinker \& Wetzel (2010). In this paper we are interested in separating these two steps
and understanding the errors that come from the halo predictions for clustering.

\subsection{Mass function fits}
\label{massfuncsection}

\begin{figure*}
\center
\includegraphics[width=130mm]{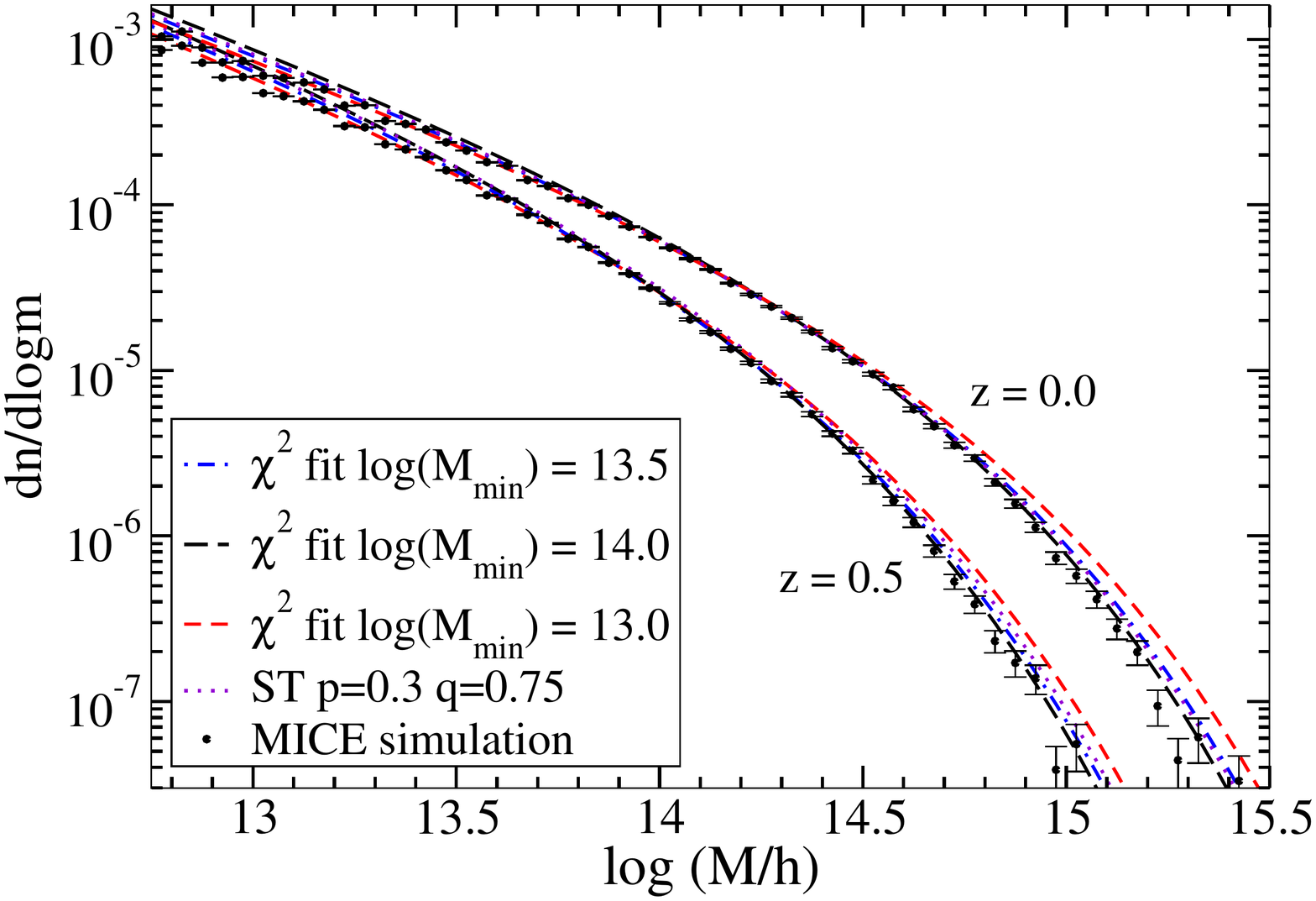}
\includegraphics[width=130mm]{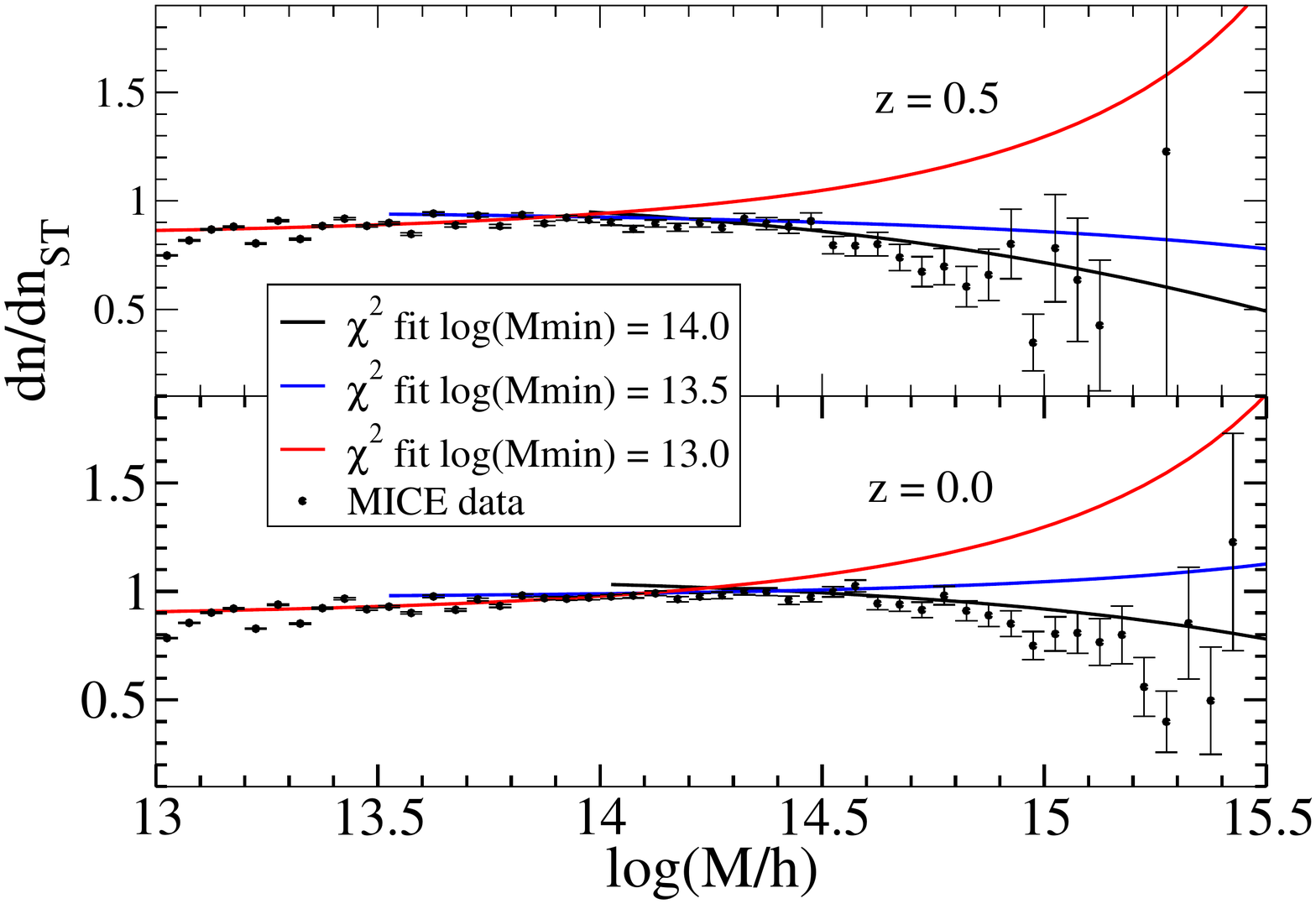}
\caption[Halo mass function from MICE simulation ]
{\textbf{TOP:} Mass function for halos in the MICE simulation at redshifts $z=0$ (upper curves)
and $z=0.5$ lower curves. Compare them with ST best fits starting at $log(M)=$ 13.0,13.5 and 14.0.
\textbf{BOTTOM:} Ratios of the MICE mass function fits and data respect Sheth and Thormen mass function with
$p=0.3$ and $q=0.75$.}
\label{figmassf} 
\end{figure*}     

We have computed the mass function of halos for the MICE simulation. We show it in Fig.\ref{figmassf}. 
Halos have been found using a Friends of Friends algorithm with a linking length 0.168 times 
the mean interparticle distance, and their masses have been corrected for discreteness effect
following Warren et al. (2006), i.e., the mass of the halo have been set equal to $M_p\, N (1-N^{-0.6})$ 
where N is the number of particles and $M_p$ the particle mass (which is $2.34 \; 10^{11} M_\odot$ in our simulation).  
The Warren correction has been experimentally set using a linking length equal to 0.2 times
while we are using 0.168. Differences in the correction, however, are very likely
to be minimal if not negligible for the halo mass range in which we
fit the mass function.  Notice also that
Crocce et al. (2009) has tested this correction for MICE simulations by means of randomly removing a fraction of
the dark matter particles as a way of lowering the mass resolution, and found it to work quite well.   

We have performed a $\chi^2$ fit to the mass function data, starting from different lower mass thresholds for halos. 
Best fits for Sheth and Tormen (ST) functional form are shown as dashed colored lines in Fig. \ref{figmassf} 
(top panel), while data is in black dots. The fits are dominated by the lower mass bins which have smaller errorbars. 
For comparison we have added a line showing the mass function with the commonly used ST fiducial 
values $(p,q)=(0.3,0.75)$.  To appreciate better the differences between fits we show, in the bottom panel,
the ratio of the best fit curves to that of the ST fiducial case. 

The values for p and q of each fit and their statistical errors are shown in the Table 1. 
Errors come from jack-knife subsampling and are computed in the following way. 
We divide our simulation in 64 compact regions with equal volume. 
Then we create a set of $J=64$ jack-knife subsamples of the data by removing each time one of these regions
from the whole sample. For each jack-knife subsample we compute the mass function and fit its
$(p,q)$ parameters. Errors are then obtained as

\begin{equation}
(\Delta \theta )^2 = \frac{(J-1)}{J}  \sum_{j=1}^J (\bar{\theta}-\theta_j)^2 
\label{jkeq}
\end{equation}
where $\theta$ is a generic name of any of the parameters in which we are interested, in this case $p$ and $q$,
and $\bar{\theta}$ is the average of $\theta$ over the jack-knife subsamples. And for the best fit values of 
our parameters we take $\bar{\theta}$. When doing the $\chi^2$ fit, errors in the mass functions are taken 
to be Poisson but results do not change significantly if they are estimated by the jack-knife method as well. 
Similar results are obtained if we divide the simulation in 27 jack-knife regions instead of 64. As it is 
shown in Table 1 we find that jack-knife errors on of $p$ and $q$ are smaller than the systematic errors that
we are trying to asses by setting different halo mass thresholds. 
\\
\\
\begin{tabular}{c c c c c c }
$log(M_{min})$ & z & p & q &  $\sigma_p$ & $\sigma_q$ \\ 
\hline  
13.0 & 0.0 & 0.334 & 0.665 & 0.001 & 0.003 \\
13.5 & 0.0 & 0.309 & 0.733 & 0.002 & 0.004 \\
14.0 & 0.0 & 0.275 & 0.786 & 0.004 & 0.006 \\
13.0 & 0.5 & 0.347 & 0.691 & 0.001 & 0.003 \\
13.5 & 0.5 & 0.312 & 0.763 & 0.003 & 0.004 \\
14.0 & 0.5 & 0.280 & 0.801 & 0.010 & 0.011 \\
\end{tabular}
\newline
{\bf Table 1} Best fit values of the Sheth and Tormen's $p$ and $q$ parameters to the simulation mass function,  
and their jack-knife errors $\sigma_p$ and $\sigma_q$.

\subsection{Comparison with the local model}
\label{compsec}
 
We compare the bias predictions from the mass function fits with the measured local bias from scatter plots,
in Fig. \ref{b1c2mass}. Both $b_1$ and $c_2$ are shown as a function of halo mass, 
and for both redshifts that we are studying. 
We find that, generically, the predicted linear bias $b_1$ falls below the local bias. 
This happens for all three mass thresholds we use to fit the mass function. 
\footnote{The only exceptions are halos above $7 \cdot 10^{13} M_{\sun}$ at $z=0.5$, 
where predictions seems to be above measurements.
This is because at these masses convergence in the biasing parameters
as a function of $R_s$ (i.e., in Fig.\ref{figb1Rs}) 
has not been reached for $R_s=30$. For this halos, we have 
checked that if we use a higher smoothing radius (i.e, $R_s=60$ Mpc/h)
we recover the general trend where $b_1$ measure in the scatter plot
is above ones from the mass function.}

The best agreement between the linear local bias and the predictions is when the mass function is fitted for
masses above $10^{14} M_{\sun}$. The lower the mass threshold to fit the mass function the worst the agreement
between measurements and predictions. For a threshold of $M > 10^{13}
M_{\sun}$ predictions are completely
misplaced, for a threshold of $M > 10^{13.5} M_{\sun}$ we have differences of about 5-10\%, while if the
threshold is $M > 10^{14} M_{\sun}$ differences are of few percent. This few percent agreement however 
have to be taken with caution because we are using the high mass halo tail to 
predict the bias of a halo sample in which most of the halos had not contributed to the ST fit. 

In the same Fig. \ref{b1c2mass}, for comparison with most ST plots in the literature, 
we have shown also the predictions for the fiducial ST case of $p=0.3$ and $q=0.75$. 
Its performance is similar to the one with a threshold of $10^{13.5} M_{\sun}$,
i.e, with differences about 5-10\% with the local bias measurement. 
If we where to use the values $p=0.3$ and $q=0.707$ that also exist in the literature 
it would yield much lower values of $b_1$, thus we confirm the convenience of using higher
values for $q$ as suggested in Sheth and Tormen (2001).

We show the $c_2$ values from the scatter plot (black dots with
errorbars) in the bottom plot of Fig. \ref{b1c2mass},
against the ST predictions from the mass function fit (dashed lines). As expected 
$c_2$ errors from the scatter plot are larger than errors in $b_1$ since it is more difficult to fit
the second order of the Taylor expansion than the first one. 
For the predictions, statistical errors have been computed using Jack-Knife subsampling 
(also for $b_1$) but they are not shown because they are much smaller
than the systematics we see by changing the mass threshold.

\begin{figure}
\center
\includegraphics[width=88mm]{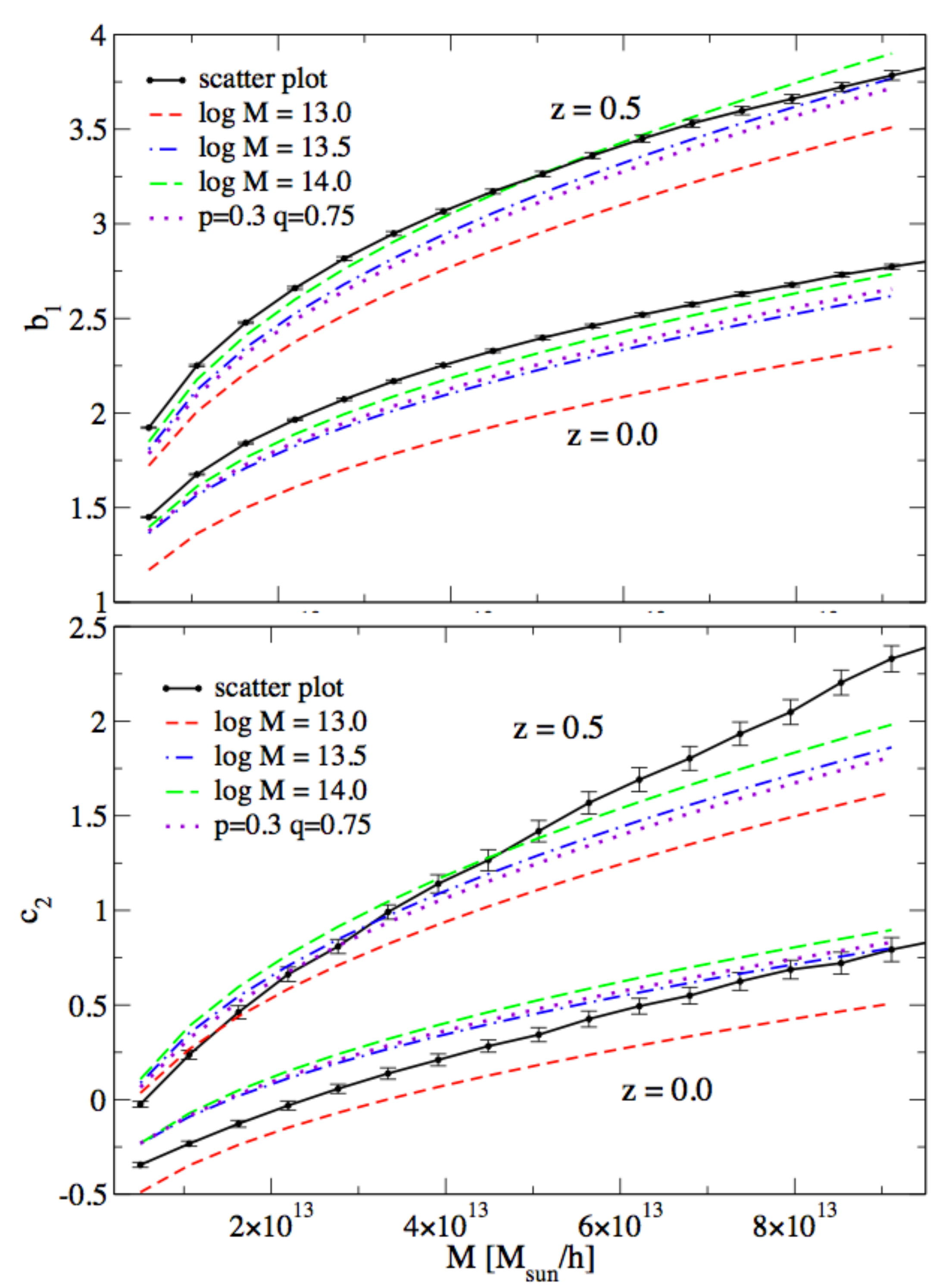}
\caption[Dependence of $b_1$ and $c_2$ on the halo mass]
{Variation of $b_1$ (top panel) and $c_2$ (bottom panel) as a function
of the halos mass. In black we show the values measured directly from
the $\delta_h$-$\delta_m$ local relation in the simulation with a
smoothing of $R=30Mpc/h$ and compare them with   
ST predictions (dashed lines) and the fiducial ST p=0.3 q=0.75 case (dots).
As labeled in the figure each panel have both $z=0$ and $z=0.5$ results.}
\label{b1c2mass}
\end{figure}

\subsection{Comparison with clustering}

So far we have compared the scatter plot bias values both with bias from clustering
statistics (section \ref{sec:localbias}) and with ST predictions (this section).
This comparisons have allowed the study of the local bias model. Since the local
bias is not a direct observable in observations we now proceed to compare directly
the bias predictions from the mass function with the bias from clustering.
This comparison for $b_1$ and $c_2$ is shown in Fig. \ref{b1clusteringtoST}. 
For reference we have also included the fiducial ST prediction with $p=0.3$ and $q=0.75$

We find that the clustering of both $b_{hm}$ and $b_{hh}$ are slightly
higher than the ST predictions. Recall that we have shown that the Poisson shot-noise correction 
does not work for $b_{hh}$. The correct shot-noise correction
is smaller, see Eq.\ref{eq:shot-noise}, and produces values of
$b_{hh}$ that are close to $b_{hm}$ 
Thus the apparent agreement between $b_{hh}$ and the mass
function predictions for $z=0$ is just a fluke and one should only
compare to $b_{hm}$ which is not affected by shot-noise. The values of
$c_2$ are also affected by the shot-noise correction.

Similar differences between predictions and measurements where found by Manera et al. (2010) 
when studying the large scale bias from other set of simulations. 
If we are looking only at the $b_{hh}$, ST predictions work at 5-10\% level at $z=0.5$  
As we will comment in section \S \ref{sec:masscal} this could be enough to calibrate mass of halos 
at about the same percent level. For greater precision more  elaborate  modeling is needed.

\begin{figure}
\center 
\includegraphics[width=95mm]{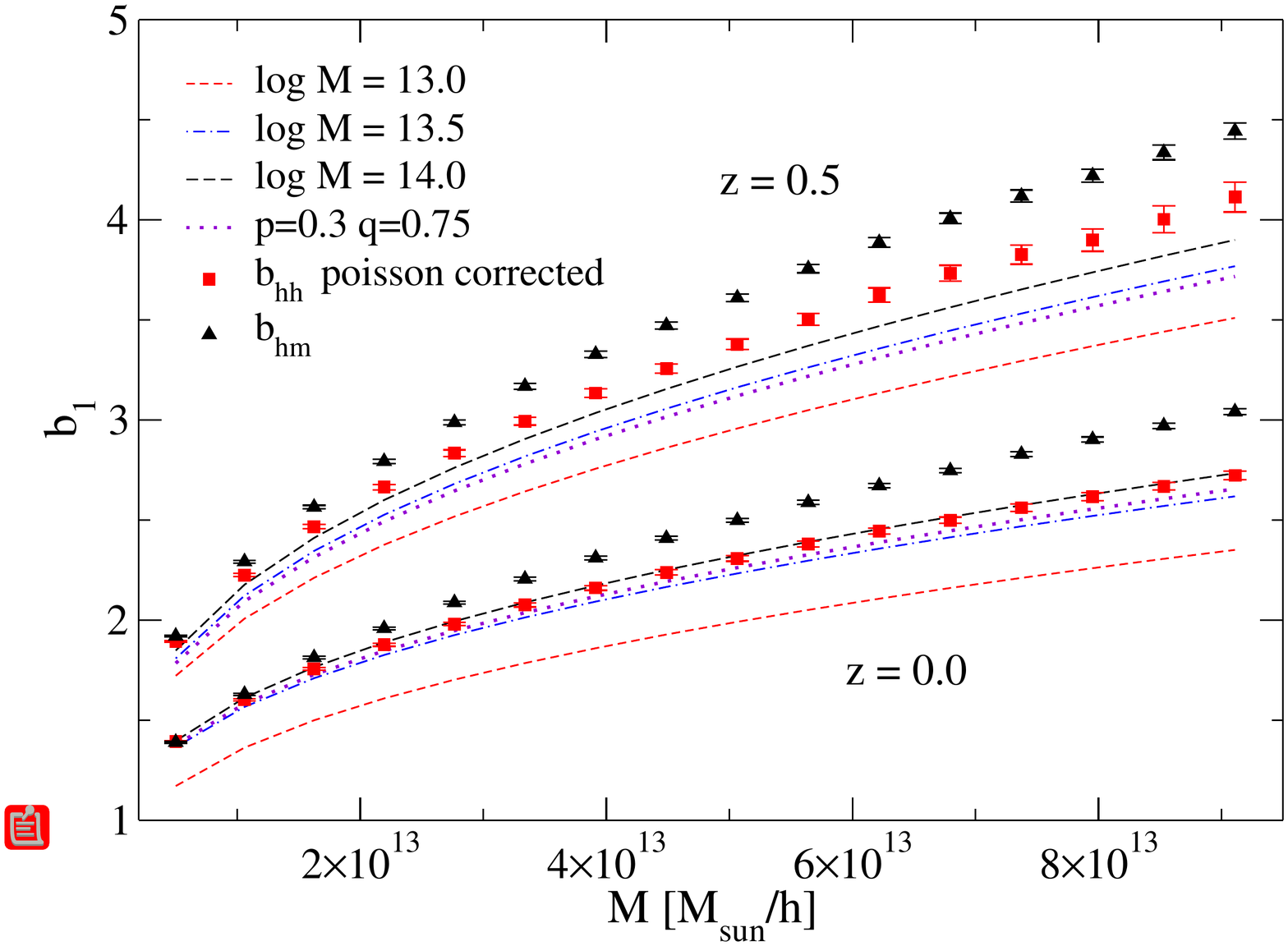}
\vskip -0.7cm
\includegraphics[width=95mm]{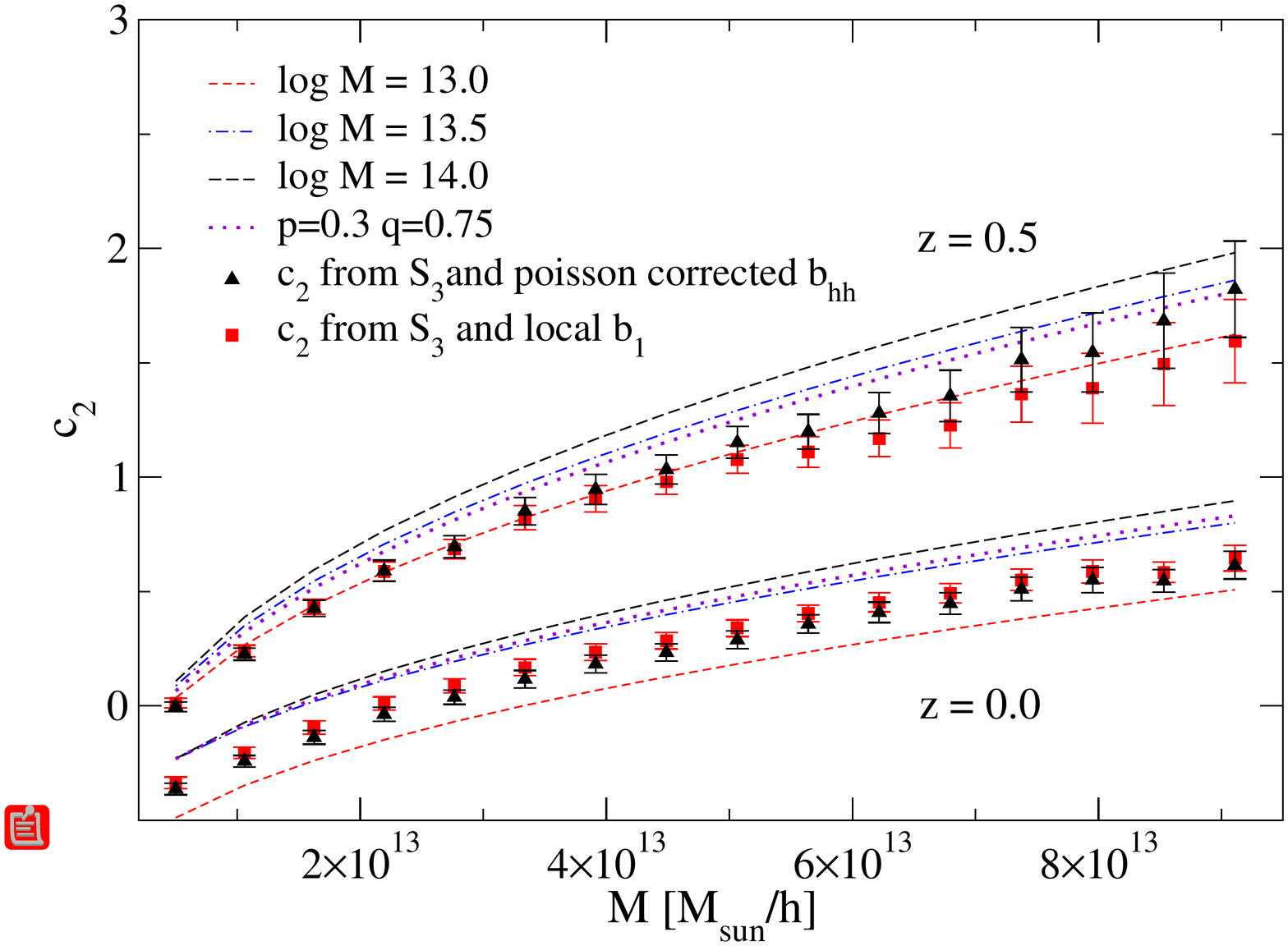}
\caption
{Comparison of the linear and second order bias from clustering with that of ST predictions from the mass function fit. 
Errors are from Jack-knife method with 64 regions. Smoothing radius of $R=30Mpc/h$}
\label{b1clusteringtoST}
\label{c2clusteringtoST}
\end{figure}


\begin{figure}
\center
\includegraphics[width=80mm]{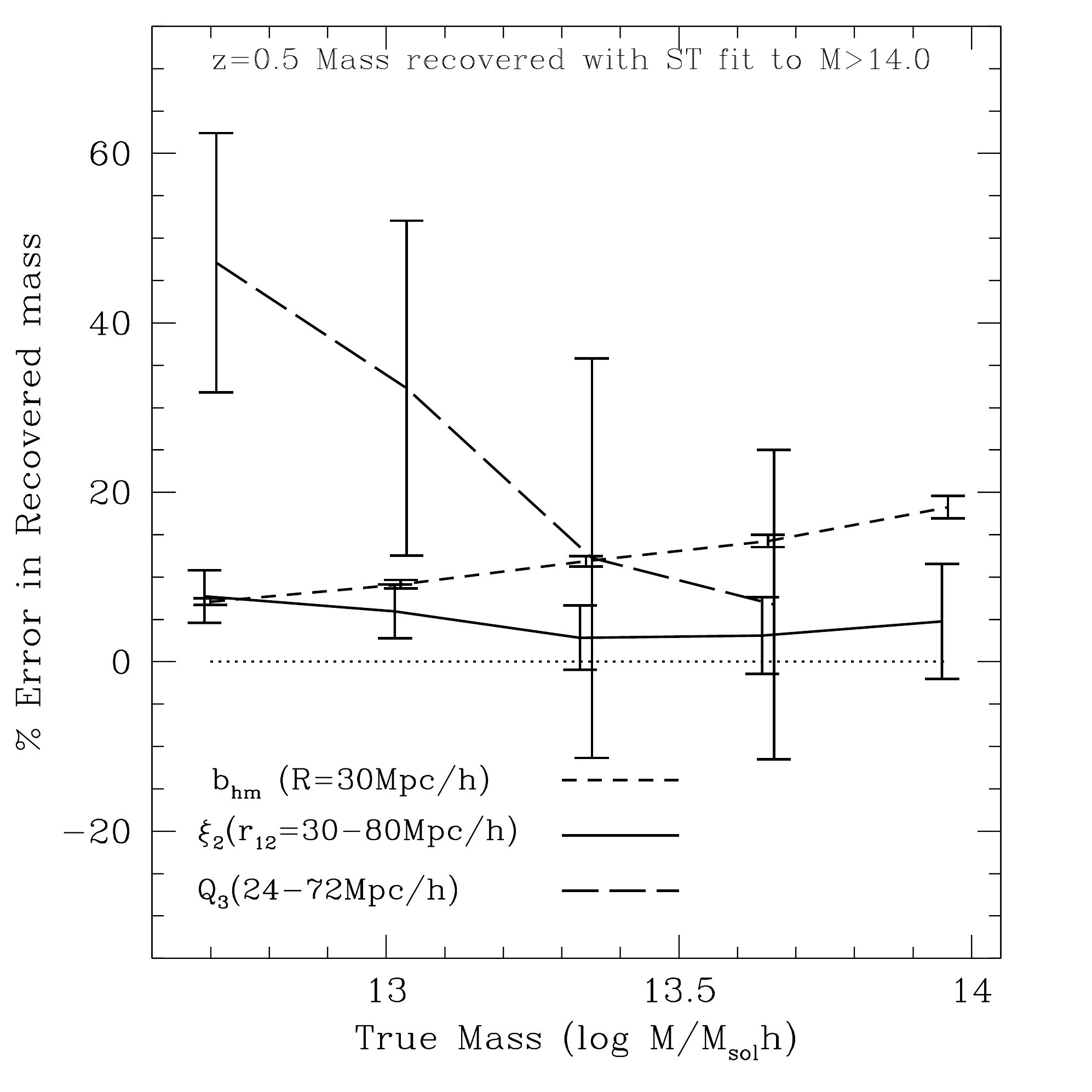}
\caption{Relative error in the Mass recovered using 
bias measurements from clustering together with the 
bias-Mass relation in the peak-background split model
obtained from ST fit to MF for  $log M>14$.
Short-dashed line
 corresponds to the bias from the cross-variance $b_{hm}$ in section
\S 2.5. Continuous line correspond to the bias from the 2-point function 
on large 30-80 Mpc/h scales.
Long dashed line come from $Q_3$ in section \S2.3.}
\label{recovM}
\end{figure}

\subsection{Halo mass estimation}
\label{sec:masscal}

We now explore the potential use of linear bias $b$ measurements to calibrate the mass threshold
of a halos sample. For a given mass-bias ($b$-$M$) relation in the
halo model (i.e., Eq.\ref{biasparam}) we can use the measurements of bias $b$ in the halo sample 
to predict the corresponding mass threshold. This is 
illustrated in Fig.\ref{recovM}.  We have used the clustering
biases measured in the 2-point function (i.e., Fig.\ref{fig:b1x2}) and in the
variance (i.e., Fig.\ref{b1massclustering}) at z=0.5 and use the mass-bias relation  
from the the ST fit to $log M>14$ (which seems to provide the best fit to data) 
to calibrate the mass from bias. For the variance we use $b_{hm}$ (the
halo-mass cross variance) rather than $b_{hh}$ to avoid discreteness effects.

The idea of recovering the mass function from bias and variance measurements, and 
subsequently fit for cosmological parameters, have been explored by
Lima and Hu (2004, 2005, 2007) 
in the so-called self calibration method.
They assume a peak background split prediction to relate the bias to
the mass function (in
particular Eq.\ref{biasparam} with the fiducial ST values), and allow for an scatter
relation between the proxy of the mass (eg. X-ray Temperature) and the true mass. 
Fig. \ref{recovM} clearly shows that there is a bias in the recovered mass,
which will propagate into the cosmological fits as a systematic error. 
This bias in the recovered mass, could in principle be corrected with
the use of mock  samples or the non-linear corrections presented above.

To measure the bias based on the 2-point statistics we need to know $\sigma_8$. Otherwise
we just recover the value of $M $ in units of $\sigma_8$. This is not the case for $Q_3$,
which provides $M$ with independence of $\sigma_8$, but at the expense
of a larger errorbar and more systematic effects for small masses.
For large masses there are too few halos to have a reliable measurement of $Q_3$. 
Also note that observations are in redshift space while here we have only show results in
real space. We expect differences in redshift space and we defer this to future studies.

\section{Conclusions}

In this paper we have used a cosmological dark matter simulation of volume $V=(1536 Mpc/h)^3$ 
from the MICE simulation team to study the halo clustering and 
bias of halos above $2\cdot 10^{13} M_{\odot}$.  We have focused in
clustering in configuration space (as oppose to Fourier space): the
2 and 3-point correlation function, the variance and skewness and the
halo-mass cross-correlations. Our main results can be summarized as follows:

\begin{itemize}

\item
We have looked at the local deterministic biasing prescription, 
which assumes a local non-linear relation $T=f(\delta)$ between mass fluctuations, 
$\delta$, and its tracer,$T$. In simulations this relation is an approximation with significant
scatter around the mean $f(\delta)$ relation. We have fitted
this scattered relation with a parabola and found the linear bias
$b_1$ and the quadratic bias, $b_2$  
(or equivalently $c_2=b_2/b_1$) at different smoothing scales (see Figs 1 \& 2). 
We show that constant biasing values are reached for smoothings larger
than $30-60Mpc/h$. This provides a new interpretation for the so-called local model:
it local only on average over very large scales. This has an immediate application
for bias calculations as one can set $R_s\rightarrow\infty$ in practice
and neglect many of the next to leading order terms in a multipoint
expansion.

\item
We have measured the correlation function of halos, $\xi_h(r)$,
and compared it to the matter correlation function, finding that the bias
is  approximately constant at scales larger than 20Mpc/h (see Figs 2-3) as predicted by
the local bias model.
Given our errors, there is some room for a small (few percent) scale dependence at
scales near the BAO. 
We have shown (see Fig. \ref{fig:b1x2})
that this bias in the correlation is very well matched by the local
bias  prediction from the scattered $\delta_m$-$\delta_h$ parabola, 
when we use a large smoothing $R_s$ (where convergence to constant
values is reached)

\item
We have measured the 3-point correlation function of halos and fitted its shape to obtain
$b_1$ and $c_2 = b_2/b_1$. We have shown (see Fig. \ref{fig:b1q3}) that the linear bias
obtained from the 3-point correlation function 
 does not quite match the bias of the 2-pt correlation function 
(or the local bias, which are the same within errors) at our lower mass bins:
$M < 10^{13} M_\odot/h\simeq$ 50 particles. The 3-point predictions
for $b_1$ follow well the qualitative behavior of bias as a function of
mass and redshift but there are some systematic (2-sigma) deviations
at the lower mass end with good agreement for $M > 10^{13} M_\odot/h$.
For $M > 10^{14} M_\odot/h \simeq$ 500 particles, 
error bars start becoming too large to conclude.  

\item 
We have measured the bias from the halo cross-variance $b_{hm}=<\delta_h \delta>/<\delta^2>$ 
and found that it differs from the local bias at about 10\%, or even more for the
most massive halos (see Fig. \ref{b1massclustering}). 
The true local bias can be recovered,
if we include non-linear correction (using the measured $b_2$ of the local model). 
This is in contrast to the bias from the 2-pt correlation function for which there 
was no need of including nonlinear terms. 

\item 
We have measured the bias from the halo variance $b^2_{hh}=(<\delta_h^2>-1/\bar{n})/<\delta^2>$  
and found that it is different from $b_{hm}$ and from the linear local bias. We have shown that
in order to be able to predict $b_{hh}$ from the local bias we need to take into account \textit{both} 
nonlinear and stochastic effects (see Fig. \ref{b1massclustering}). 

\item 
We have shown that the appropriate discreteness correction to the variance is sub-Poisson, 
and found that its ratio to the Poisson term $1/n$ is approximately constant in our range of masses.
(see Fig. \ref{scattermoment}). 
Overcorrecting the variance with $1/n$ masks the nonlinear contributions, 
thus giving an estimated value of $b_1$ apparently closer to that of
the local bias (specially at z=0), 
as we show in Fig. \ref{b1massclustering}. 

\item
We have fitted the mass function of halos with a Sheth and Thormen functional form
and applied the peak-background split Ansatz to predict the bias parameters. 
These predictions depend significantly on the mass threshold used to
fit the mass function and they give systematically lower linear bias (about 5-10\%)  than that
measured in clustering or local relation  (see Fig. \ref{b1clusteringtoST}
and Fig.\ref{b1c2mass}).

\item
Finally, we have estimated the mass of halos from the measured bias (Fig. \ref{recovM}), 
showing that there is a systematic error when using the common ST peak-background split prediction. 
These systematic errors have to be taken into account when recovering the mass function from clustering of
halos, since they will propagate to the estimator of cosmological parameters, like the
dark energy equation of state.

\end{itemize}

We can conclude from the above that the different bias predictions are
only accurate to 5-10\% level. 
In the case of the 2-point functions (auto and cross-correlations), 
the local model seems accurate and we find that the origin of the discrepancy
lies in the peak-background prescription. This is not so clear for the
3-point function, where probably both assumptions contribute to the error.
For the smoothed moments, we find that next to leading order and discreetness
corrections (to the local model) are needed at the $10-20\%$ level.
Although this accuracy might still be adequate for current
data, where typical errors are 10-20\% 
(eg. Norberg et al. 2002, Zehavi et al. 2005, Gaztanaga et al. 2005,
Nichol et al 2006), more work needs to be done to narrow this to the percent
level that will be likely needed in upcoming and  future surveys for precision
cosmology and better understanding of galaxy evolution.


\section*{Acknowledgments}  

We thank all the MICE collaboration team, and specially P. Fosalba
who provided the central positions of halos in the simulation.
We thank Roman Scoccimarro for his comments on the first draft of
this paper.
The MICE simulations have been developed at the MareNostrum
supercomputer (BSC-CNS) thanks to grants AECT-2006-2-0011 through
AECT-2010-1-0007. Data products have been stored at the Port
d'Informació Científica (PIC). 
This work was partially supported by NSF AST-0607747,
NASA NNG06GH21G and NSF AST-0908241, and 
the Spanish Ministerio de Ciencia e Innovacion (MICINN), projects
AYA2009-13936, Consolider-Ingenio CSD2007- 00060 and
research project 2009-SGR-1398 from Generalitat de Catalunya.

\end{document}